\documentclass{aa}  

\usepackage{graphicx}
\usepackage{txfonts}
\usepackage{natbib}
\usepackage{xcolor}
\bibpunct{(}{)}{;}{a}{}{,}
\usepackage{xcolor}
\usepackage{appendix}
\usepackage[colorlinks=true,citecolor=blue,linkcolor=blue]{hyperref}
\usepackage{threeparttable}
\usepackage{amsmath}
\usepackage{bm}
\begin{document}

\title{Spectral properties of the neutron star low-mass X-ray binary 4U 1636--53, XTE J1739--285 and MAXI J1816--195}

\author{Zhenyan\ Fei  \inst{1,2}
\and
Ming\ Lyu \inst{1,2}
 \and 
Guobao \ Zhang \inst{3,4}
\and
Xuejuan\ Yang  \inst{1,2}
\and
Federico Garc\'{\i}a \inst{5}
}

\institute{Department of Physics, Xiangtan University, Xiangtan, Hunan 411105, China
\and
Key Laboratory of Stars and Interstellar Medium, Xiangtan University, Xiangtan, Hunan 411105, China\\
\email{lvming@xtu.edu.cn}
 \and
 Yunnan Observatories, Chinese Academy of Sciences (CAS), Kunming 650216, P.R. China
 \and
 Key Laboratory for the Structure and Evolution of Celestial Objects, CAS, Kunming 650216, P.R. China 
  \and
Instituto Argentino de Radioastronom\'ia (CCT La Plata, CONICET; CICPBA; UNLP), C.C.5, (1894) Villa Elisa, Argentina
 }

\abstract
{

We investigated simultaneous NICER plus NuSTAR observations of three neutron star low-mass X-ray binary 4U 1636--53, XTE J1739--285 and MAXI J1816--195 using the latest reflection models, with the seed photons feeding into the corona originating from either the neutron star (NS) or the accretion disk. We found that, for the sources in the hard spectral state, more than $\sim$ 50\% of the NS photons enter into the corona if NS provides seed photons, while only $\sim$ 3\%-5\% disk photons go to the corona if seed photons come from the disk. This finding, together with the derived small height of the corona, favors the lamp-post geometry or boundary layer scenario where the corona is close to the central neutron star. Additionally, we found that the source of the seed photons has big influence in the significance of the NS radiation, especially for the soft spectral state. This result may help explain why the NS radiation in MAXI J1816--195 is weak in the previous work. More importantly, for the first time, we explored the properties of the corona in the NS systems with the compactness ($l-\theta$) diagram. We found that the corona in the NS systems all lie in the left side of the pair-production forbidden region, away from the predicted pair-production lines. This finding indicates that either the corona in these NS systems is not pair-dominated, possibly due to the additional cooling from NS photons, or the corona is composed of both thermal and non-thermal electrons.

}
 
{}
\keywords{X-rays: binaries; stars: neutron; accretion, accretion disc; X-rays}
\titlerunning{Spectroscopy of 4U 1636--53, XTE J1739--285 and MAXI J1816--195}

\maketitle


\section{Introduction}  

X-ray binaries are systems that are composed of a compact object and a main sequence or post main sequence companion star. If the companion star mass is less than 1 M$_{\odot}$, then the system is classified as a low-mass X-ray binary (LMXB), which could be further divided into a black-hole (BH) LMXB or a neutron star (NS) LMXB according to the nature of the central compact object. In a LMXB, the compact object accretes matter from the companion via Roche lobe overflow \citep{tauris06}, with vast amount of the gravitational potential energy released in X-rays.

There exist some differences between the two types of the LMXBs. The most obvious one is the presence of the physical surface in a NS LMXB, which could lead to significant observation features like thermonuclear bursts or mHz QPOs due to nuclear burning of the accreted material accumulated on the surface \citep[e.g.,][]{lewin93,revni01,stroh06,heger07,lyu15}, or different spectral evolution behavior \citep{done03}, or pulsation originated from the residual magnetic field \citep{done07}. Besides, for a NS system, there is a so-called boundary or spreading layer \citep{ss86,Inoga99} between the innermost part of the accretion disk and the surface of the neutron star. In-falling matter decelerates from the Keplerian velocity at the inner accretion flow to the surface rotating velocity of the NS in this region, with its kinetic energy deposited.

Radiation scenario in a NS LMXB is more complicated than in a BH LMXB. In LMXBs, there is usually a multi-color thermal emission from the accretion disk and a non-thermal power-law radiation from a corona composed of high-energy plasma. In addition, the power-law photons could illuminate the disk and hence generate a reflection component. There exists, however, additional thermal radiation from the NS surface in a NS system, and this thermal emission could also illuminate and then reflect off the accretion disk. Furthermore, the thermal photons coming from the NS surface could enter into the corona and be up-scattered as power-law photons. 

Observationally, spectral properties of the NS LMXBs have been widely investigated, however, the role of the NS radiation is still unclear. On the one hand, the influence of the NS illumination to the disk is poorly understood. Most of the work only explore the scenario where the disk is illuminated and ionised by only the corona, while studies focusing in the NS illumination is limited \citep[e.g.,][]{cackett10,wilkins18,ludlam20,ludlam22,Garcia22,lyu23}; On the other hand, due to possible degeneracy problems \citep[e.g.,][]{sanna13,lyu14}, the scenario where the seed photons coming from the NS surface for the Compton scattering has not been fully explored in the current stage.  

In order to get more insight into the influence of the neutron star radiation, in this work we performed an investigation into three NS LMXBs which have been observed simultaneously by the Neutron Star Interior Composition Explorer (NICER) detector and the Nuclear Spectroscopic Telescope Array (NuSTAR) telescope. Here we give a brief introduction of these three sources as below:


4U 1636--53: It harbors a millisecond pulsar with a spin frequency of 581 Hz \citep{zhang97,stroh02}. The orbital period is 3.8 hr \citep{van1990-1636-3.8h}, at a distance of 6.0 $\pm$ 0.5 kpc \citep{Galloway2006-1636-6kpc}. This source shows variety of observational features like thermonuclear bursts, super-bursts, mHz QPOs, kHz QPOs and time lags. Besides, 4U 1636--53 shows evolution of all the spectra states \citep{belloni07,altamirano08}, with a regular state transition cycle $\sim$ 40 days \citep{shih05,belloni07}.

XTE J1739--285: \citet{Galloway2008-Dis} derived the distance of XTE J1739--285 to be less than 7.3 kpc based on peak flux of photospheric radius-expansion (PRE) bursts. \citet{Bailer-Jones-2018-4u1735} estimated the distance of 4 kpc with parallaxes published in the $Gaia$ data. Fe K$\alpha$ line and Compton hump have been detected in the spectrum of this source \citep{mondal22}. \citet{Kaaret-1739-1122burst} reported a 1122 Hz burst oscillation in XTE J1739--285, indicating that it may be the only neutron star discovered to rotate within a sub-millisecond period. But later studies have not detected the existence of this special burst oscillation \citep{Galloway2008-Dis, Bilous2019-no1122Hz, Bult2021ApJ-1739-no1122}. \cite{Bult2021ApJ-1739-no1122} found that a 386.5 Hz oscillation was the more prominent signal and believe that the rotation period of XTE J1739--285 may not be a sub-millisecond period.

MAXI J1816--195: It is a new accreting millisecond X-ray pulsar (AMXP) discovered by Monitor of All-sky X-ray Image (MAXI) Gas Slit Camera (GSC) on June 7th, 2022 \citep{Negoro2022A-1816transient}. Its circular orbital period is $\sim$ 4.8 hours \citep{Bult2022ApJ-J1816}, and the spin frequency of MAXI J1816--195 is 528 Hz \citep{Bult2022-528hz}. \citet{Chen2022-6.3kpc} derived the upper limit of its distance to be 6.3 kpc from the peak flux of the brightest burst detected by the Hard X-ray Modulation Telescope (Insight-HXMT) from 2022 June 8 to 30, and \citet{Bult2022ApJ-J1816} obtained the upper limit as 8.6 kpc from the peak flux of bursts with NICER. Besides, \citet{Lizs23} reports the detection of X-ray pulsations in this source up to $\sim$ 95 keV.

\section{Observations and data reduction}

In this work, we analyzed the simultaneous NICER plus NuSTAR observations of 4U 1636--53, XTE J1739--285 and MAXI J1816--195. These observations (Table \ref{OBS}) were taken from April 2019 to June 2022, with the total exposure time being $\sim$ 26.4 ks and $\sim$ 153 ks for the NICER and the NuSTAR observations, respectively. We used HEASOFT ver 6.30.1 to reduce the data from the two satellites.

\subsection{NICER observations}
NICER is located on the International Space Station (ISS), with the X-ray Timing Instrument (XTI) operating in the soft X-ray range of 0.2-12 keV \citep{Gendreau16}. The energy resolution is $\sim$ 85 eV at 1 keV and $\sim$ 137 eV at 6 keV.\footnote{https://heasarc.gsfc.nasa.gov/docs/nicer/nicer$\_$about.html} We performed the standard data reduction using the NICER Data Analysis Software using the latest calibration files ver 20240206. The clean XTI events are extracted using the command $nicerl2$, with standard calibration applied. Time intervals containing X-ray bursts were removed by the tool $xselect$. We obtained the background by running the tool $nibackgen3C50$ \citep{Remillard22}, and used command $nicerrmf$ and $nicerarf$ for the production of the response matrix files (RMFs) and the ancillary response files (ARFs). Finally, we applied the command $grppha$ to group the spectra to ensure a minimum of 100 counts per bin.

\subsection{NuSTAR observations}
NuSTAR has two independent detectors FPMA and FPMB operating in a broad energy range of 3-79 keV, with a spectral energy resolution of 400 eV (FWHM) at 10 keV \citep{harrison13}. The data from the two detectors are calibrated and screened to get clean events with the pipeline tool $nupipeline$ using the latest calibration files ver 20240916. We applied the script $nuproducts$ for the extraction of the source and the background spectra using a circular region of radius 150 arcsec. The response and ancillary response files were produced simultaneously with this script. The spectra were rebinned with the default option $grpmincounts=100$ in $nuproducts$.
\\
\\

\begin{figure*}[!htpb]
\centering
\includegraphics[scale=0.18]{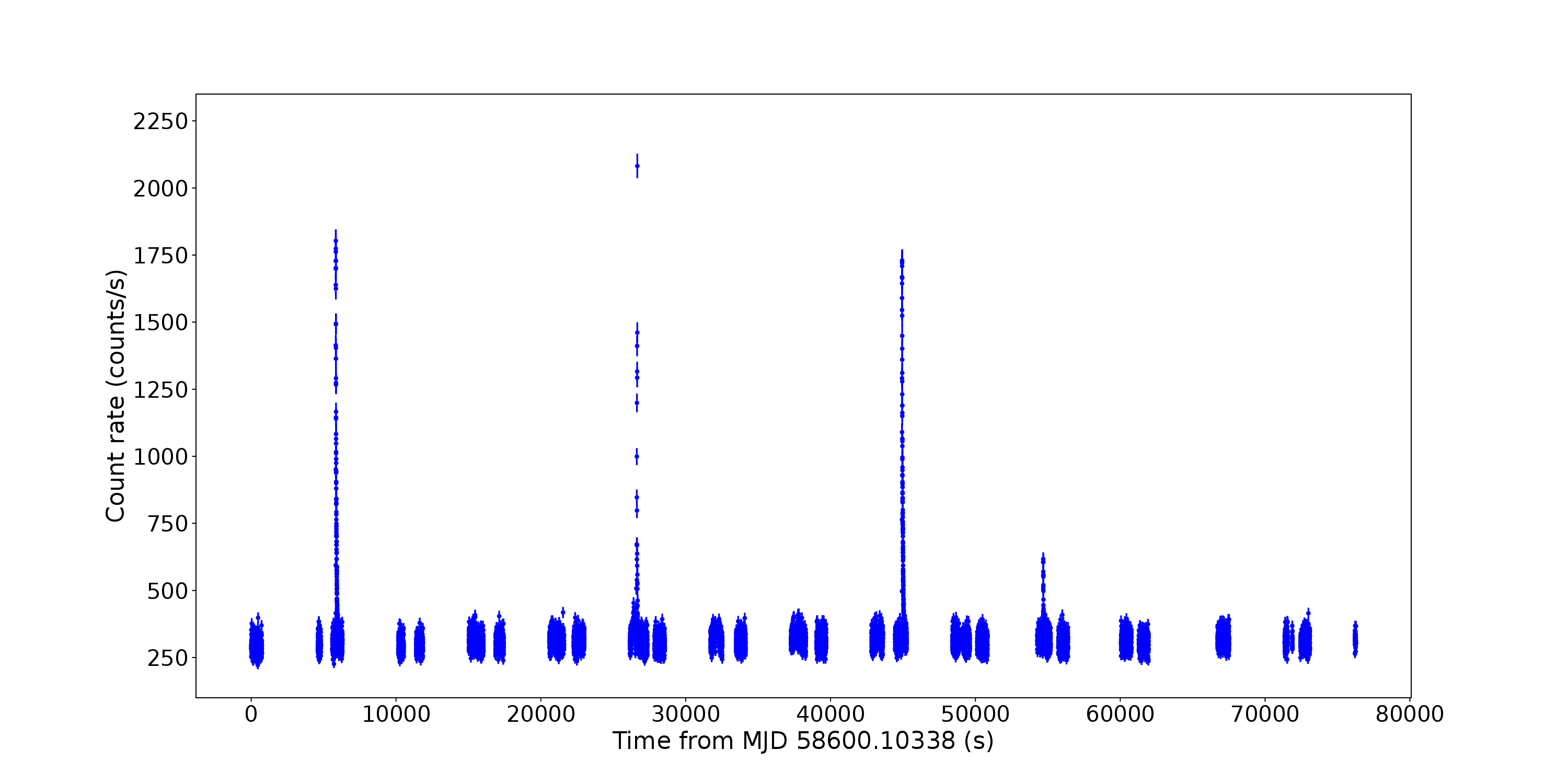}
\includegraphics[scale=0.18]{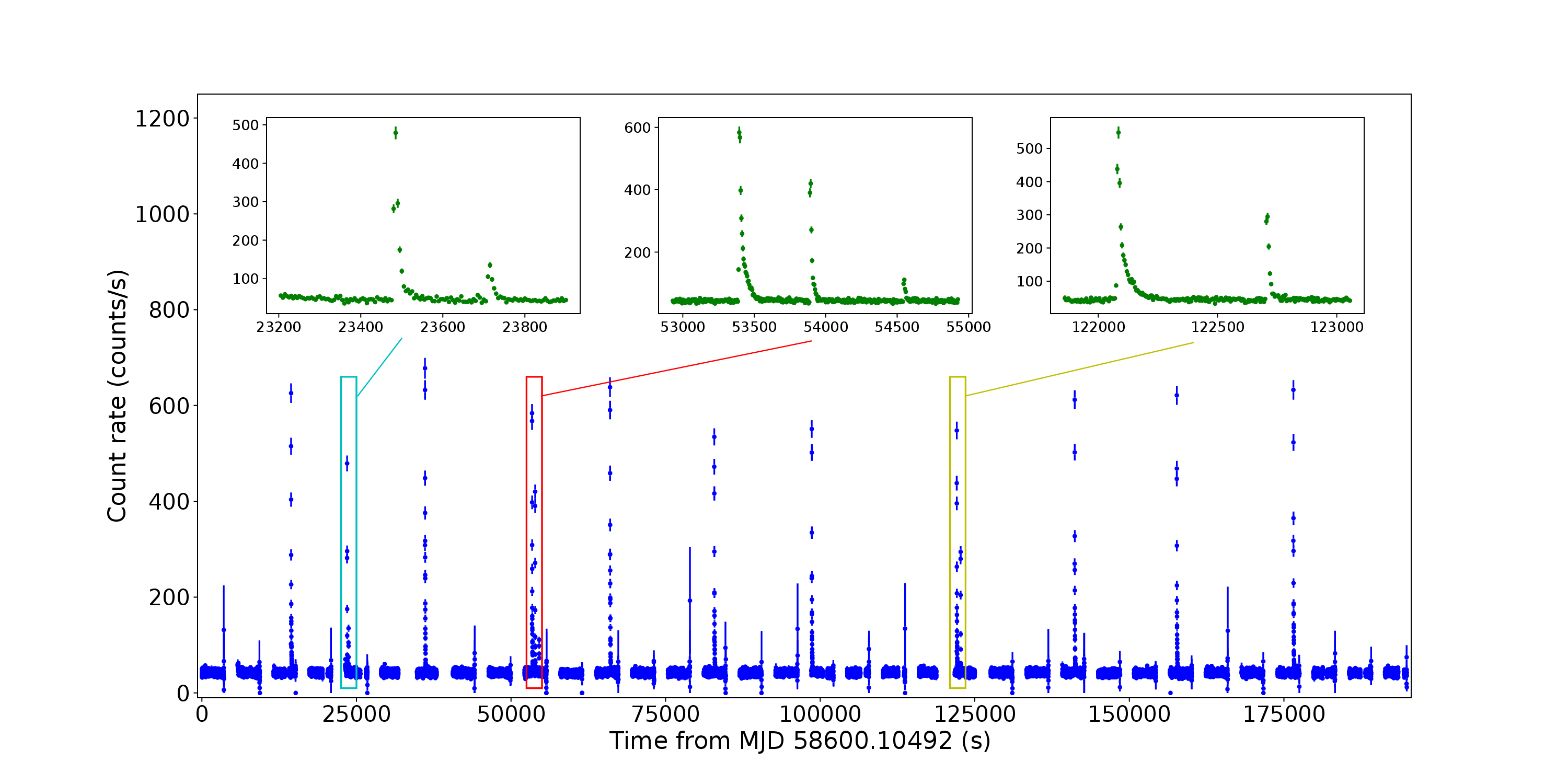}
\includegraphics[scale=0.18]{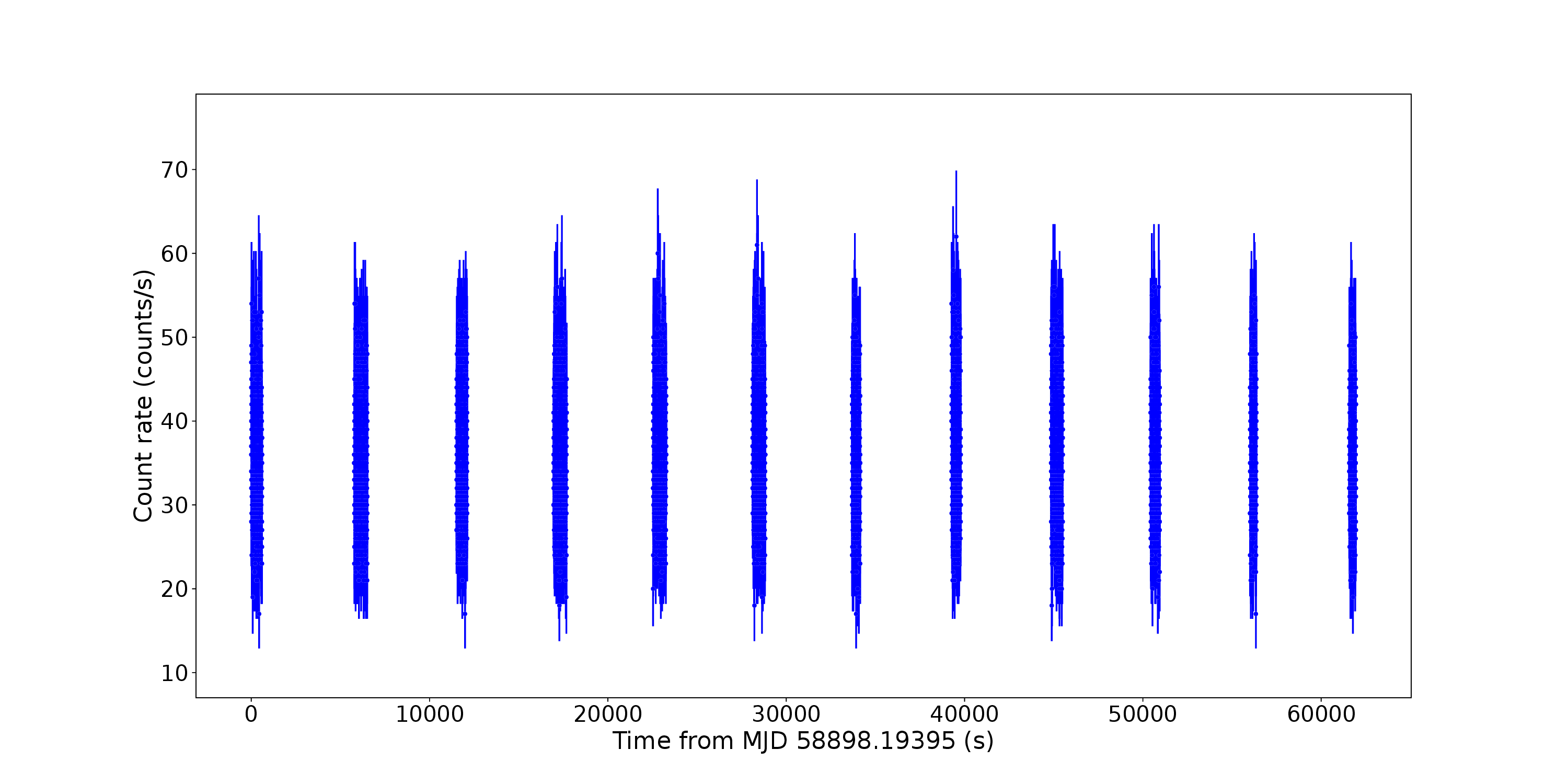}
\includegraphics[scale=0.18]{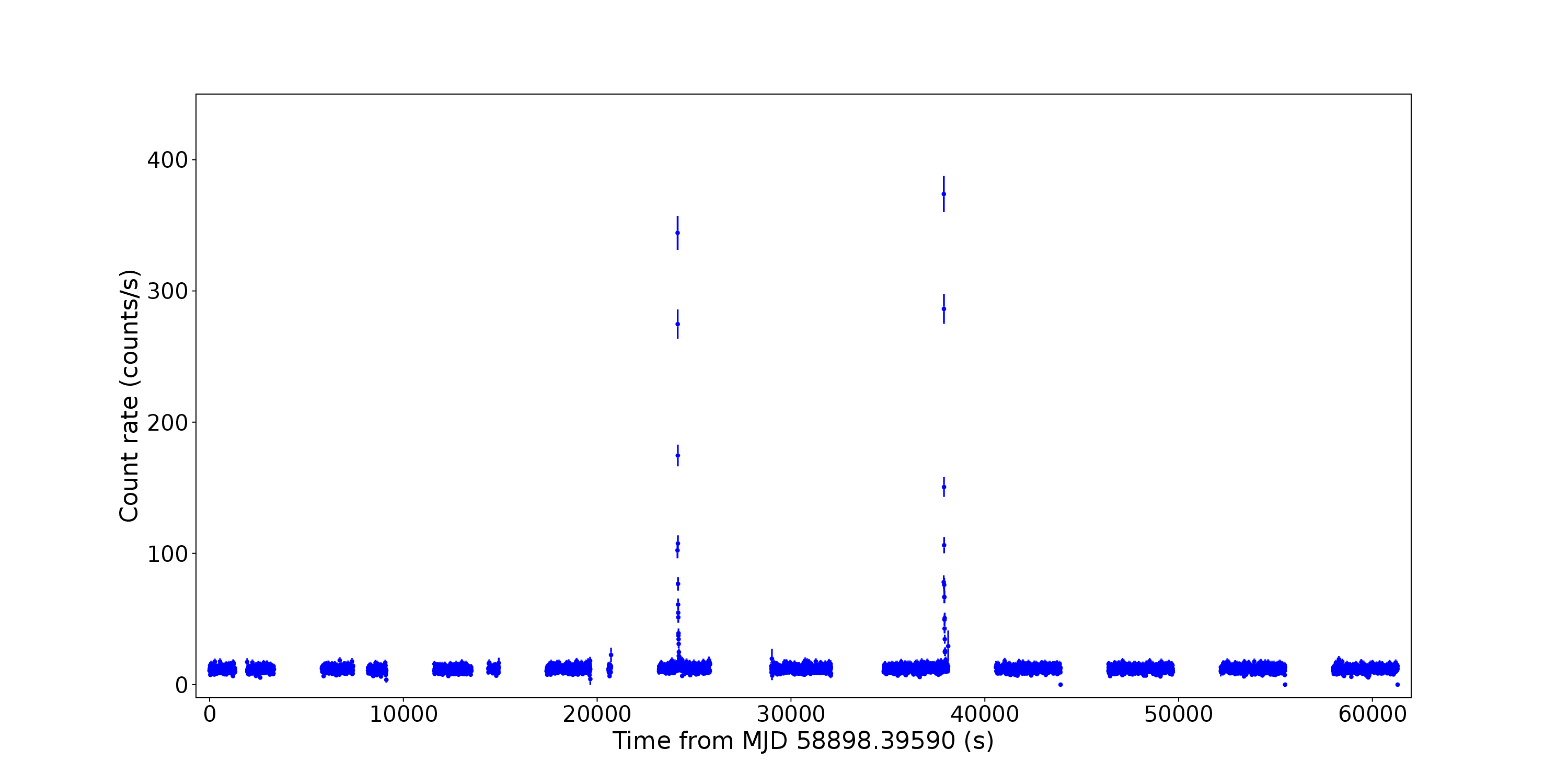}
\includegraphics[scale=0.18]{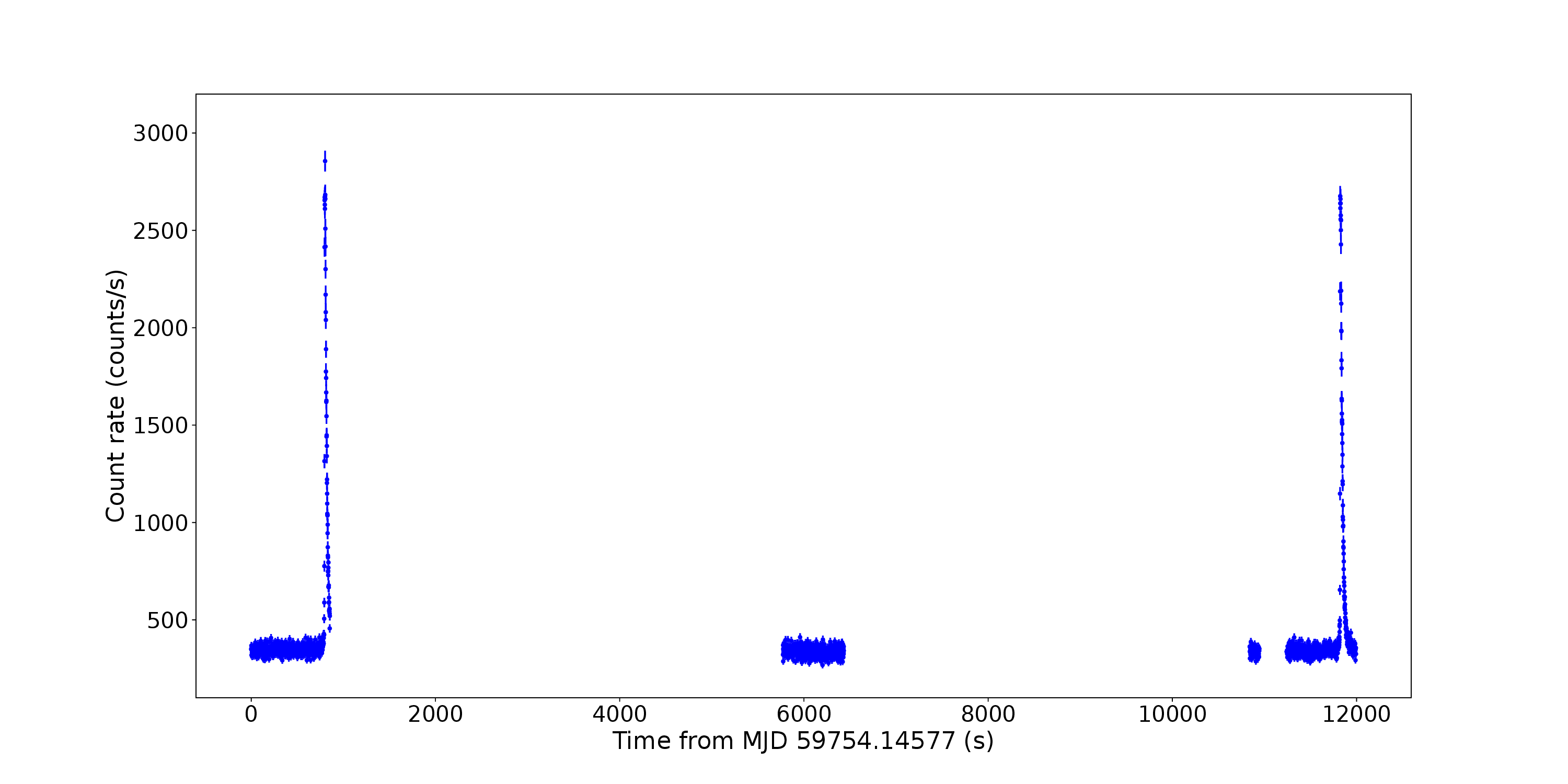}
\includegraphics[scale=0.18]{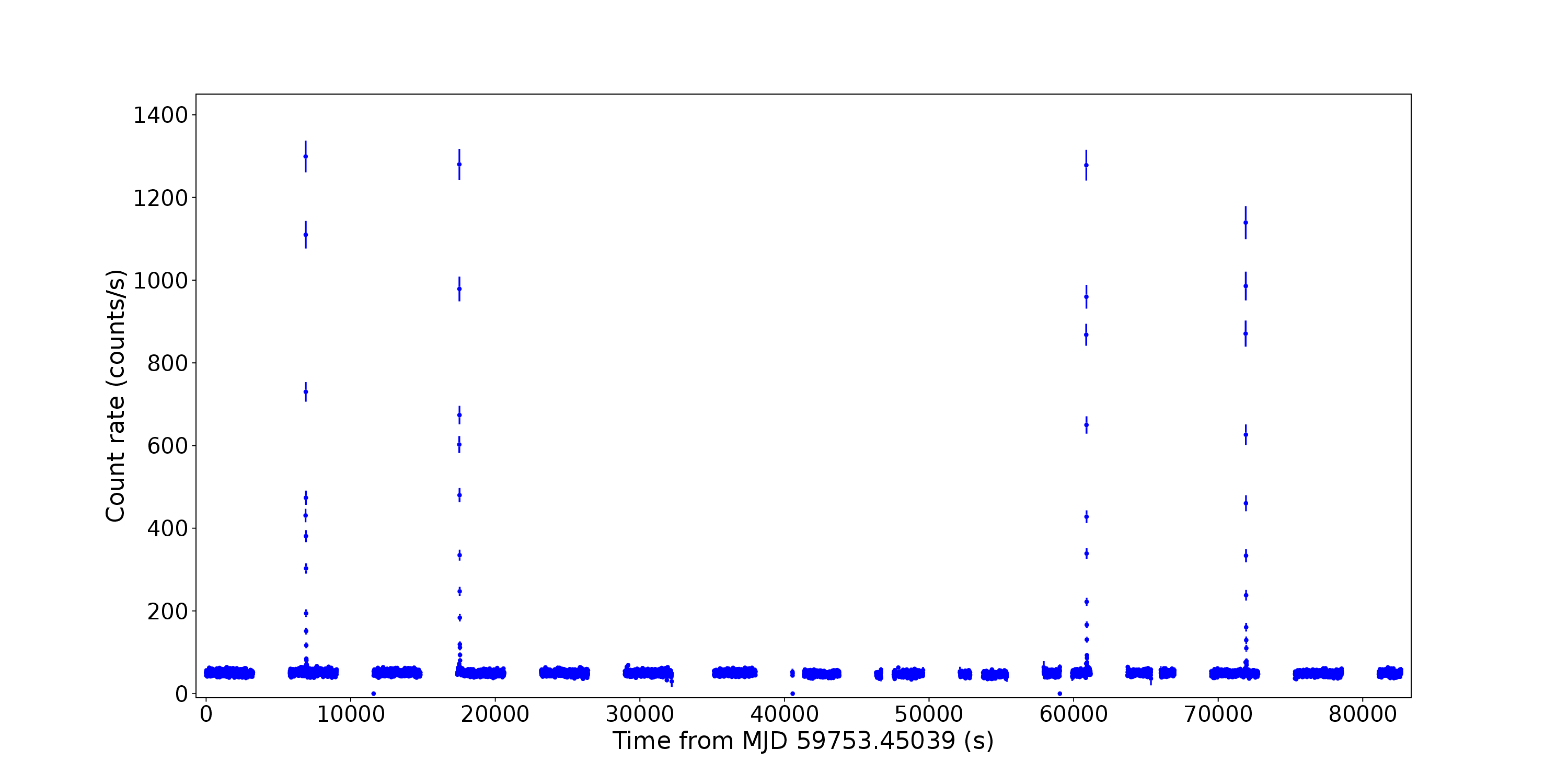}

\caption{NICER (left panel) and NuSTAR/FPMA (right panel) light curve of 4U 1636--53 (top), XTE J1739--285 (middle) and MAXI J1816--195 (bottom). We extracted NICER light curves in 0.5-10 keV at a time resolution of 1 second, and derived the NuSTAR light curves in 3-79 keV using a time bin of 5 second. }
\label{light-curve}
\end{figure*}

As shown in Figure \ref{light-curve}, there are 4 and 15 bursts in the NICER and the NuSTAR observation of 4U 1636--53. For MAXI J1816--195, there are 2 bursts in the NICER data and 4 bursts in the NuSTAR data. The number of bursts in the NICER and the NuSTAR observation of XTE J1739--285 is 0 and 2, respectively.

\begin{figure*}[!htpb]
\centering
\includegraphics[scale=0.23]{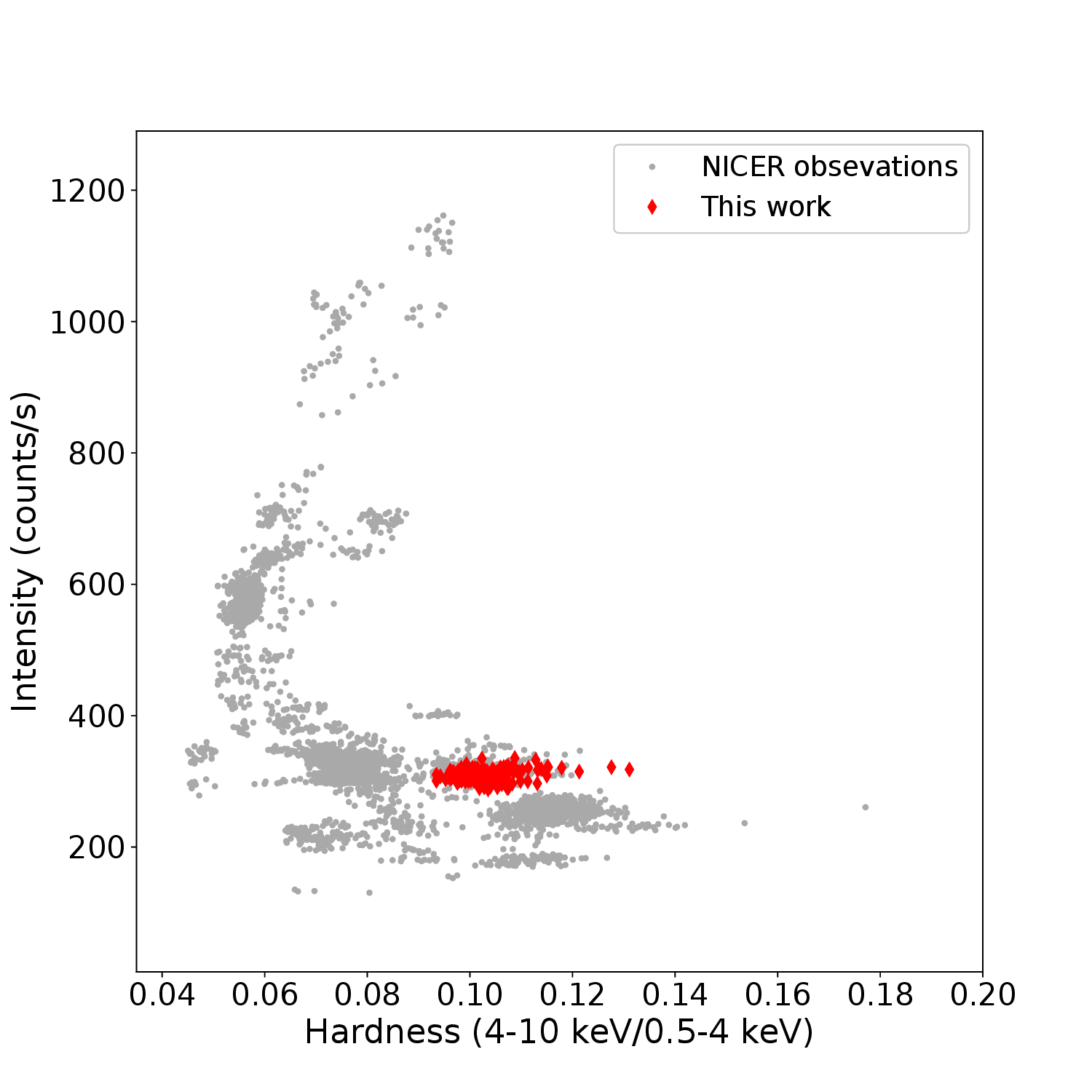}
\includegraphics[scale=0.23]{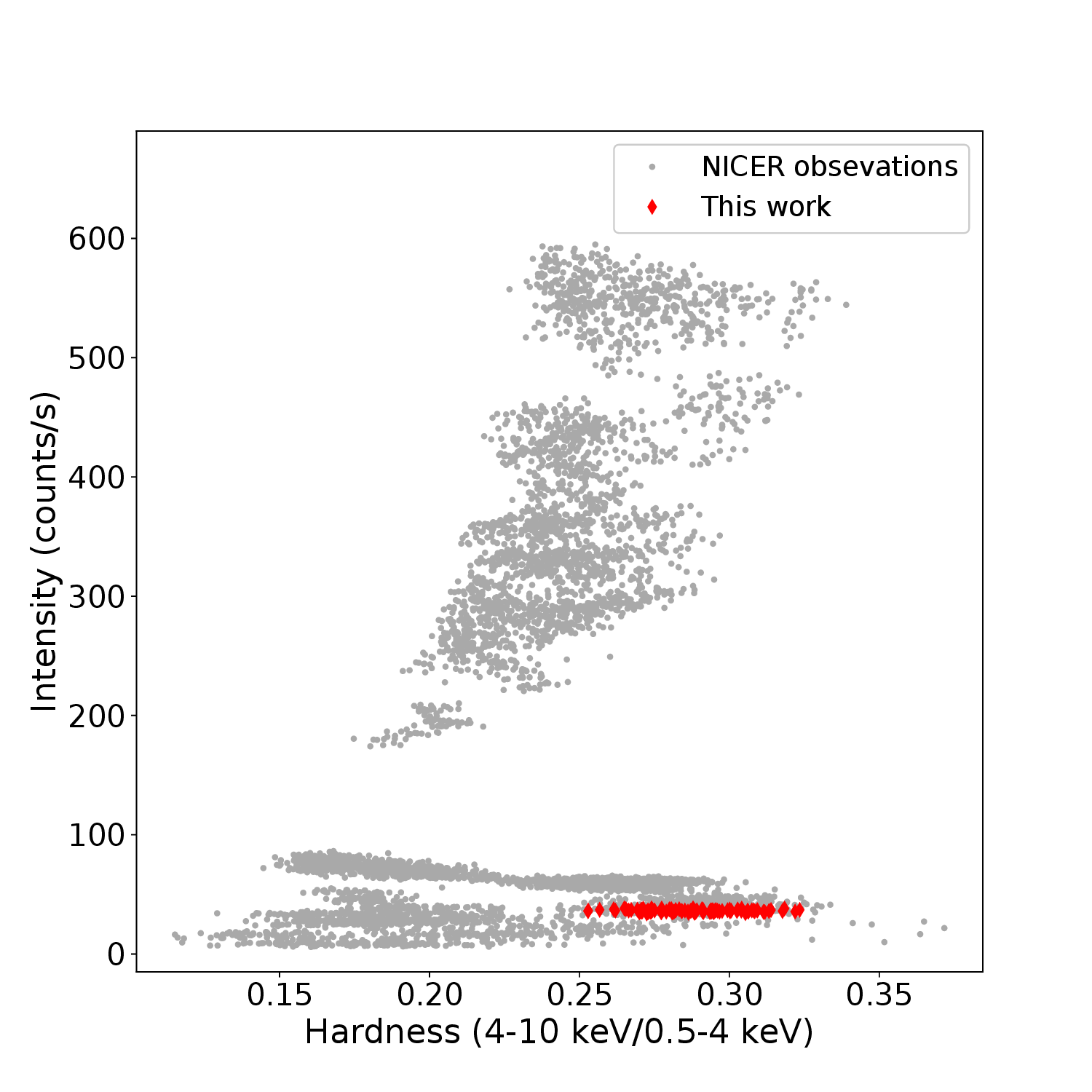}
\includegraphics[scale=0.23]{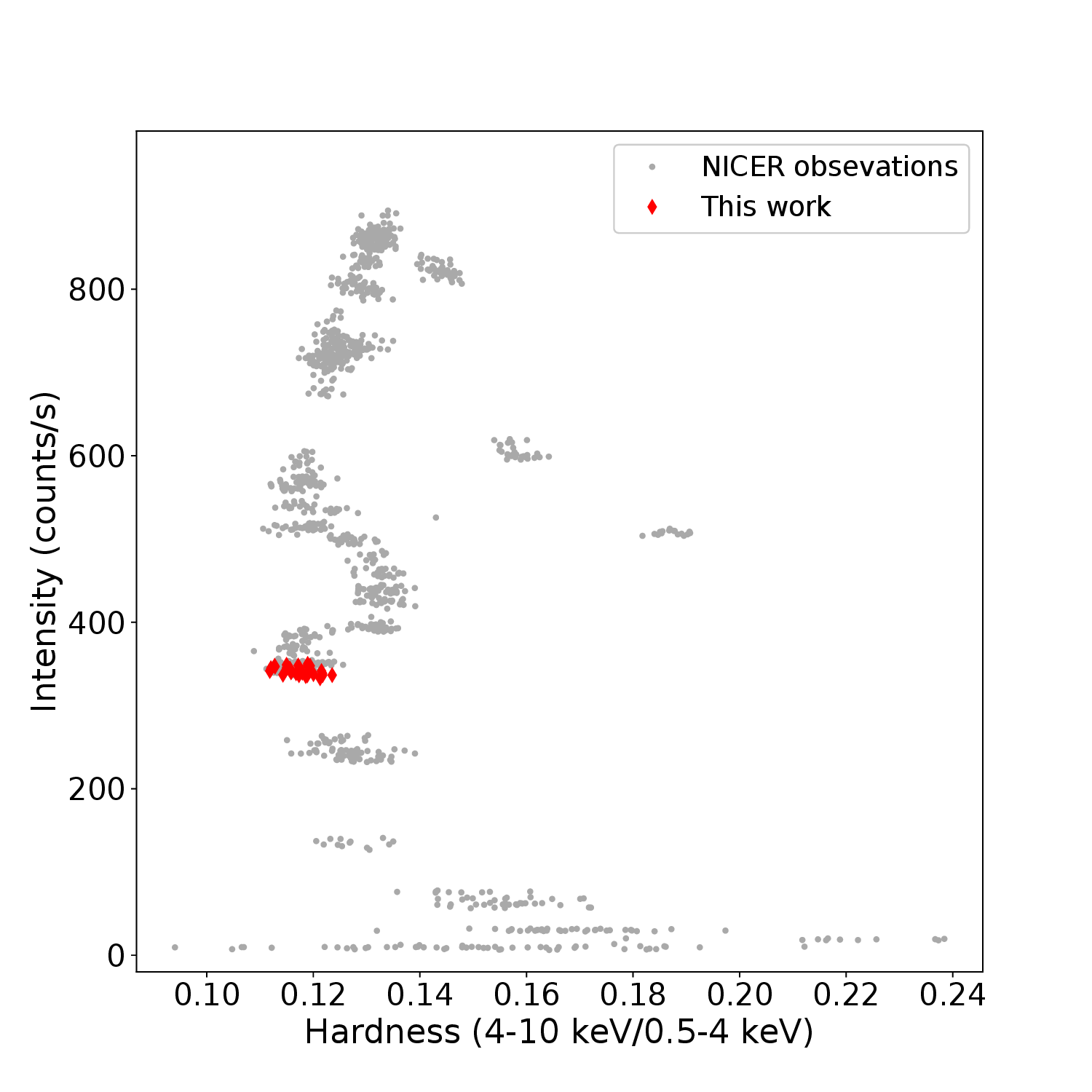}
\caption{Hardness-intensity diagram of NICER observations of 4U 1636--53 (left), XTE J1739--285 (middle) and MAXI J1816--195 (right). The hardness ratio is defined as the count ratio between the 4-10 and 0.5-4 keV, and the intensity is defined as the counts in 0.5-10 keV. Each datapoint in the plots corresponds to a 64-second NICER data segment, with the NICER observations analyzed in this work shown in red and other NICER observations marked in gray.
}
\label{HID-ALL}
\end{figure*}

In order to study the spectral state of the sources, we further generated the hardness-intensity diagram (HID) (Figure \ref{HID-ALL}) using the NICER observations. We first extracted light curves in 0.5-4 keV, 4-10 keV and 0.5-10 keV band every 64 seconds. And then we calculated the hardness as the count rate ratio between the 4-10 keV and the 0.5-4 keV, and defined the intensity as the count rate in 0.5-10 keV. We found that the spectra of 4U 1636--53 and XTE J1739--285 used in this work were in hard spectral state, while MAXI J1816--195 was in a much softer state. 
	
\begin{table*}
\centering
\caption{NICER/NuSTAR observations of the three NS LMXBs in this work.}
\begin{tabular}{ccccc}
\hline
Source Name       &Telescope  &  Observation ID        &Start time                &Exposure time (ks) $^{*}$\\
		
4U 1636--53      &NICER       &2050080211              & 2019-04-27 02:26:16      &$\sim17.2$  \\
                 &NuSTAR     &30401014002            & 2019-04-27 02:26:09        &$\sim88.2$  (FPMA);  $\sim88.9$ (FPMB) \\
\hline       		      
XTE J1739--285    &NICER      &2050280129             & 2020-02-19 04:33:40      &$\sim7.0$    \\
	              &NuSTAR     &90601307002            & 2020-02-19 09:16:09      &$\sim28.7$ (FPMA);  $\sim28.8$ (FPMB) \\
\hline 
MAXI J1816--195   &NICER      &5533011601             & 2022-06-24 03:18:19      &$\sim2.2$ \\
	              &NuSTAR     &90801315001            & 2022-06-23 10:26:09      &$\sim35.1$ (FPMA); $\sim35.3$ (FPMB) \\
		
\hline       				
\end{tabular}
\tiny
\\
$^{*}$ Final exposure time after excluding instrumental drops and X-ray bursts.    
\label{OBS}
\end{table*}

\begin{table}
\centering
\caption{Reflection models applied to the spectra of the three NS LXMBs in this work. We used the model M1-M4 to fit the spectra of 4U 1636--53 and XTE J1739--285 in the hard spectral state, and used the model M1-M5 to describe the spectra of MAXI J1816--195 reported in the soft state.}
\begin{tabular}{cc}
\hline
M1    &    $Tbabs\times(diskbb+thcomp \times bbodyrad+relxillCp)$  \\
M2    &    $Tbabs\times(bbodyrad+thcomp \times diskbb+relxillCp)$  \\
M3    &    $Tbabs\times(diskbb+thcomp \times bbodyrad+relxilllpCp)$ \\
M4    &    $Tbabs\times(bbodyrad+thcomp \times diskbb+relxilllpCp)$  \\
M5    &    $Tbabs\times(diskbb+compTT+relxillNS)$  \\

\hline
\end{tabular}
\label{models}
\end{table}

\section{Spectral analysis}

We used XSPEC version 12.12.1 to fit the NICER and the NuSTAR spectra together in this work. We first selected an energy range of 1.0-70 keV, with the NICER and the NuSTAR data covering 1.0-10.0 keV \citep[e.g.,][]{mondal22} and 3.0-70 keV \citep[e.g.,][]{mon19,karga23,dias24}, respectively. However, we found that there remains residuals at the low energies of the NuSTAR spectra, similar to the residuals caused from possible instrumental difference reported in the work of \citet{Madsen-2020-4keV}. We thus ignored the NuSTAR data below 4 keV. For XTE J1739--285, we used the energy only up to 60 keV as the spectra became background dominated above that. During the fits, we selected the model $tbabs$ to describe the interstellar absorption along the line of sight, using the solar abundance of \citet{Wilms2000ApJ} and the photon-ionisation cross section table of \citet{verner96}. A  $constant$ component was added to describe possible cross-calibration between different instruments.

We first tried a simple model $bbodyrad+diskbb+cutoffpl$ to estimate the spectra shape. The $bbodyrad$ component fits the thermal radiation from the neutron star surface, and the $diskbb$ component \citep{Mitsuda84,Maki86} describes multi-color blackbody emission from the accretion disk. We also included a $gauss$ component to fit the possible iron emission line in 6.4-6.97 keV. We found that the spectra could not be well described, possibly due to the lack of reflection continuum in the model. We then replaced the $cutoffpl$ component with a physical model $thcomp$ \citep{Zdziarski2020-thcomp}. The $thcomp$ is a convolution thermal-Comptonization model, which could more accurately represent the thermal Comptonization compared with the simple $cutoffpl$ component by taking into account the electron temperature $kT_{e}$ and the cover fraction parameter $f_{cov}$. We selected either the $bbodyrad$ component or the $diskbb$ component as the source of the seed photons for the $thcomp$ component. And we extended the energy by the command "energies 0.01 1000.0 1000 log" for the application of the $thcomp$.

\begin{figure*}[!htpb]
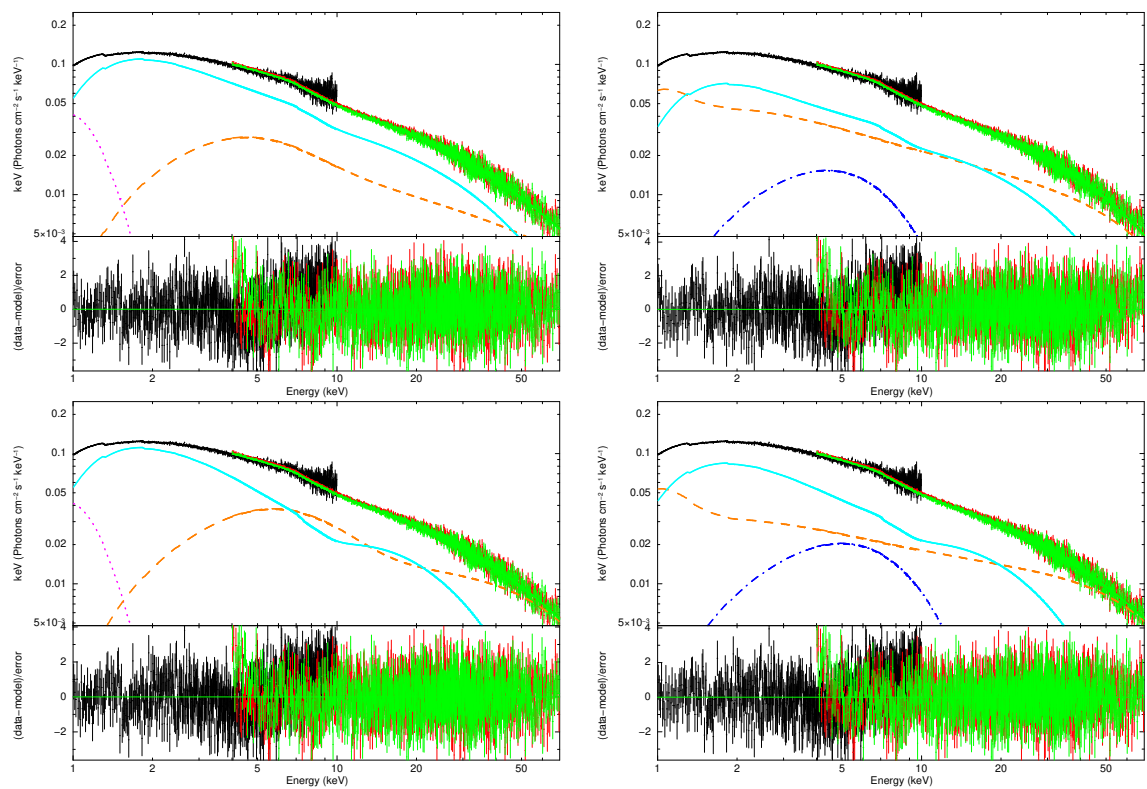

\centering         
\includegraphics[scale=0.3,angle=270]{figs/1636-4_diskbb_thcomp-bbodyrad_relxillCp_E1-10_20241012-1.eps}
\includegraphics[scale=0.3,angle=270]{figs/1636-6_bbodyrad_thcomp-diskbb_relxillCp_E1-10_20241013-1.eps}
\includegraphics[scale=0.3,angle=270]{figs/1636-5_diskbb_thcomp-bbodyrad_relxilllpCp_E1-10_20241013.eps}
\includegraphics[scale=0.3,angle=270]{figs/1636-7_bbodyrad_thcomp-diskbb_relxilllpCp_E1-10_20241012.eps}

\caption{The best fitting results with different reflection models (M1 and M2 from left to right on the top; M3 and M4 in the bottom) for NICER/NuSTAR observation of 4U 1636--53. We show the fitted spectra and the individual model components (main panel), and the residuals in terms of sigmas (sub panel) in the plots. The components $diskbb$, $blackbody$, ($thcomp \times diskbb$) or ($thcomp \times bb$), and $relxillCp$ or $relxilllpCp$ are plotted with purple dotted line, deep blue dot-dashed line, orange dashed line, light blue solid line, respectively.}
\label{dbb_lpCp_bb_1636}
\end{figure*}

\begin{figure*}[!htpb]
\centering
\includegraphics[scale=0.3,angle=270]{figs/J1739-4_diskbb_thcomp-bbodyrad_relxillCp_E1-10_20250108-3.eps}
\includegraphics[scale=0.3,angle=270]{figs/J1739-6_bbodyrad_thcomp-diskbb_relxillCp_E1-10_20250108-3.eps}
\includegraphics[scale=0.3,angle=270]{figs/J1739-5_diskbb_thcomp-bbodyrad_relxilllpCp_E1-10_20250108-1.eps}
\includegraphics[scale=0.3,angle=270]{figs/J1739-7_bbodyrad_thcomp-diskbb_relxilllpCp_E1-10_20250108-2.eps}	
\caption{The best fitting results with different reflection models (M1 and M2 from left to right on the top; M3 and M4 in the bottom) for NICER/NuSTAR observation of XTE J1739--285. We show the fitted spectra and the individual model components (main panel), and the residuals in terms of sigmas (sub panel) in the plots. The components $diskbb$, $blackbody$, ($thcomp \times diskbb$) or ($thcomp \times bb$), and $relxillCp$ or $relxilllpCp$ are plotted with purple dotted line, deep blue dot-dashed line, orange dashed line, light blue solid line, respectively.}
\label{dbb_lpCp_bb_1739}
\end{figure*}

\begin{figure*}[!htpb]
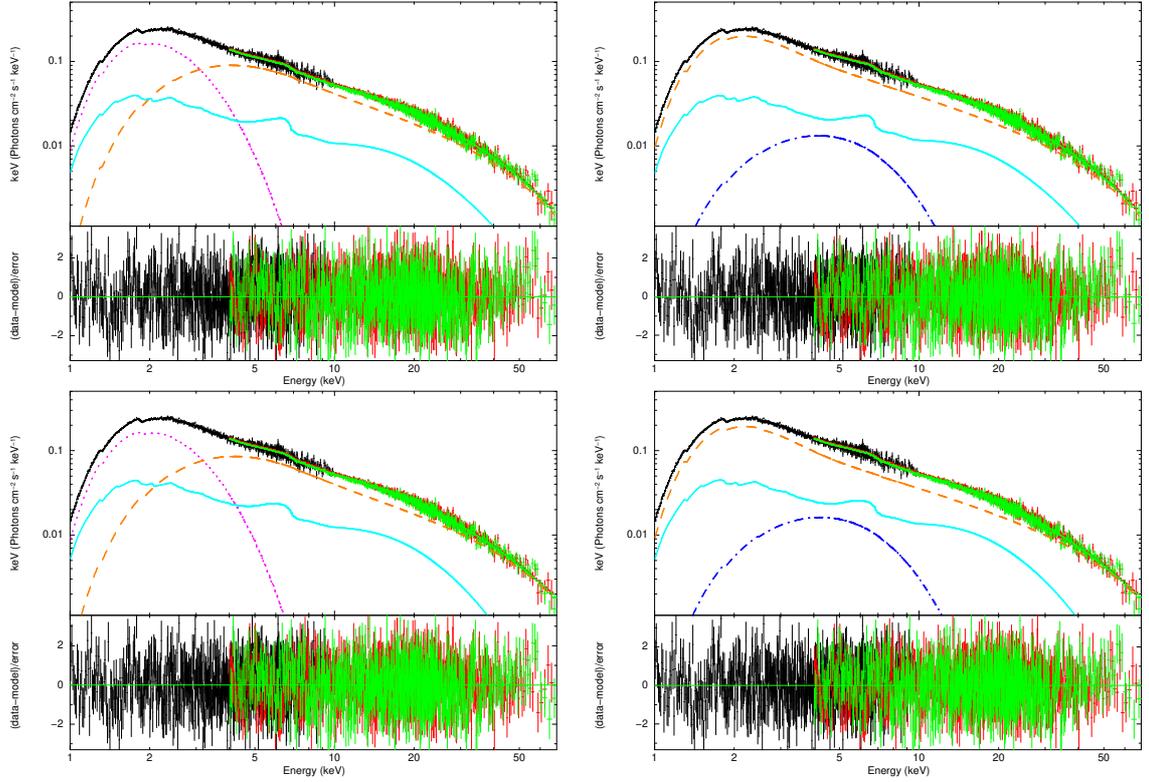

\centering
\includegraphics[scale=0.3,angle=270]{figs/J1816-4_edge-diskbb_thcomp-bbodyrad_relxillCp_E1-10_20241012-1.eps}
\includegraphics[scale=0.3,angle=270]{figs/J1816-6_edge-bbodyrad_thcomp-diskbb_relxillCp_E1-10_20241012-1.eps}
\includegraphics[scale=0.3,angle=270]{figs/J1816-5_edge-diskbb_thcomp-bbodyrad_relxilllpCp_E1-10_20241012-1.eps}
\includegraphics[scale=0.3,angle=270]{figs/J1816-7_edge-bbodyrad_thcomp-diskbb_relxilllpCp_E1-10_20241012-1.eps}
\caption{
The best fitting results with different reflection models (M1 and M2 from left to right on the top; M3 and M4 in the bottom) for NICER/NuSTAR observation of MAXI J1816--195. We show the fitted spectra and the individual model components (main panel), and the residuals in terms of sigmas (sub panel) in the plots. The components $diskbb$, $blackbody$, ($thcomp \times diskbb$) or ($thcomp \times bb$), and $relxillCp$ or $relxilllpCp$ are plotted with purple dotted line, deep blue dot-dashed line, orange dashed line, light blue solid line, respectively.}
\label{bb_lpCp_dbb_1816}
\end{figure*}

We also replaced the $gauss$ component with a more physical reflection model family $relxill$ \citep{garcia14,dauser16}. Based on the $xillver$ reflection code \citep{garcia10,garcia13} and the $relline$ ray tracing model \citep{dauser10,dauser13}, the $relxill$ models are able to calculate the reflection off the accretion disk at each emission angle. In this work, we applied the model $relxillCp$ and the model $relxilllpCp$ to fit the spectra of the three sources. The $relxillCp$ model uses a $nthcomp$ illuminating spectrum, and the $relxilllpCp$ model assumes a lamp-post geometry where the corona is a point source located at a height above the central compact object. The parameters in the model $relxillCp$ is the spin parameter, $a$, the inclination angle, $incl$, the inner and the outer radius of the disk $R_{in}$ and $R_{out}$, the emissivity index of the inner and outer disk, $q_{in}$ and $q_{out}$, the breaking radius, $R_{br}$, where the emissivity changes; the ionization parameter, $\xi$, the power law index of primary source, $\Gamma$, the electron temperature, $kT_{e}$, the disk density, $log(N)$, the iron abundance, $A_{Fe}$, the reflection fraction, $f_{refl}$, the redshift, $z$, and the normalization. For the $relxilllpCp$ model, most of the parameters are the same as in $relxillCp$, but replacing the emissivity index with two new parameters, the height of the corona, $h$, and the velocity of the primary source, $\beta$. 

In the fits, we fixed the $f_{refl}$ parameter at -1 so that only the reflection component is returned. The emissivities are fixed at the typical value of 3, and the broken radius $R_{br}$ is set to be equal to the outer radius $R_{out}$, which is fixed at 1000 $R_{g}$.The redshift is set at zero, and the disk density, $log(N)$, is fixed at 20. The spin parameter $a$ was calculated according to the formula, $a=0.47/P_{ms}$ \citep{Braje-2000-Pms}, where $P_{ms}$ is the spin period of the neutron star in milliseconds. The derived spin for 4U 1636--53, XTE J1739--285 and MAXI J1816--195 is 0.27, 0.18 and 0.25, calculated from a spin frequency of 581 Hz \citep{zhang97,stroh02}, 386 Hz \citep{Bult2021ApJ-1739-no1122} and 528 Hz \citep{Bult2022-528hz}, respectively. The primary source velocity, $\beta$, is set at zero since it is not well constrained in the fits.

For the MAXI J1816--195 data reported in the soft state \citep{mandal23}, we further applied another reflection model $diskbb+compTT+relxillns$ to explore its properties when the disk is illuminated by the NS surface instead of the corona. The $compTT$ model describes thermal Comptonization component from the corona \citep{Titarchuk1994ApJ}, with the parameters being the seed photon temperature, $T_{0}$, the plasma temperature, $kT_{e}$, the plasma optical depth, $\tau$, the geometry switch, $approx$, and the normalization. We set the $approx$ at 2 to select the sphere geometry and let the $T_{0}$ free to vary in the fit. The model $relxillns$ \citep{Garcia22} was developed to explain the reflection from the accretion disk illuminated by a blackbody component. In the $relxillns$ model, the parameters are the same as the ones in the $relxillCp$ model, except that there is no $kT_{e}$ and $\Gamma$ but instead a new parameter $kT_{BB}$. The parameter log($\xi$) is fixed at its maximum allowed value 19 and the emissivity index is set at its typical value of 3. In the fits of MAXI J1816--195, we found that there is edge-like residual near $\sim$ 1.8 keV, similar to the one reported in the work of \citet{lipp23}. This residual is proposed to be caused by the NICER calibration systematics, so we applied an additional $edge$ model to fit it.

To sum up, we used the reflection models with different illumination sources for the spectra of the three sources (Table \ref{models}); For the spectra of MAXI J1816--195 reported in the soft state, we applied an additional model with the disk being illuminated by the NS surface.

\section{Results}

In Table \ref{4u1636}, we show the best fitting results for 4U 1636--53 with the reflection model $relxillCp$ and $relxilllpCp$. We found that the temperature of the neutron star is in a range of $\sim$ 1.3-1.8 keV, and the temperature of the accretion disk is $\sim$ 0.16-0.17 keV regardless of the type of seed thermal photons. The powerlaw-index, $\Gamma$, is around 1.5-1.7, and the derived inner radius of the disk is bigger than $\sim$ 6.6 $R_{g}$, together with the normalization of the $diskbb$ component larger than 7.5$\times$10$^{4}$ in the all fits. These results are consistent with the previous reports that the source was in the hard spectral state \citep{mondal21}. The ionization parameter log($\xi$) is $\sim$ 3.6-4.0, indicating that the disk is highly ionized. The height of the corona in the lamp-post geometry is very low, $\sim$ 2.3-2.5 $R_{g}$. Interestingly, the cover fraction parameter ($f_{cov}$) in the case of NS seed photons are $\sim$ 50\%-80\%, while it is $\sim$ 5\% when the seed photons come from the disk. 

In Table \ref{j1739}, it is the fitting results for XTE J1739--285. The derived disk temperature is $\sim$ 0.14-0.15 keV, with the $N_{dbb}$ larger than 5.0 $\times$ 10$^{4}$. The temperature of the blackbody thermal emission locates in a range from $\sim$ 0.9 keV to $\sim$ 1.5 keV. The derived corona height is $2.20_{-0*}^{+0.20}$ $R_{g}$ and $2.54_{-0.17}^{+0.69}$ $R_{g}$ in the NS seed photon case and the disk seed photon case, respectively. The cover fraction is larger than $\sim$80\% in the NS seed photon case, while it is less than $\sim$ 3\% in the disk seed photon case. We show the corresponding spectra, individual components, and residuals of the fits for 4U 1636--53 and XTE J1739--285 in Figure \ref{dbb_lpCp_bb_1636} and Figure \ref{dbb_lpCp_bb_1739}.

In Table \ref{j1816}, we show the fitting results of MAXI J1816--195 with the model $relxillCp$ and $relxilllpCp$. The accretion disk is relative hot, $\sim$ 0.5-0.6 keV, with the normalization of the $diskbb$ component less than $\sim$ 2100. The blackbody temperature is $\sim$ 1-1.1 keV and $\sim$ 1.3-1.4 keV in the case of NS seed photon and disk seed photon, respectively. The photon index $\Gamma$ is $\sim$ 2, with the electron temperature $kT_{e}$ is $\sim$ 13-15 keV. The derived height of the corona is $\sim$ 3 $R_{g}$, and the disk is moderately ionized, with the ionization parameter log($\xi$) $\sim$ 2.2. The derived cover fraction parameter is $\sim$ 0.8 in the NS seed photon case, while it is relative low, $\sim$ 0.3 in the disk seed photon case. Interestingly, in both of the two models, we found that the normalization of the blackbody component changes significantly from $\sim$ 5 in the disk seed photon case to be larger than $\sim$ \text{62} in the NS seed photon case, indicating that the source of the seed photons could greatly influence the neutron star radiation. The corresponding spectra, individual components, and residuals of the fits are shown in Figure \ref{bb_lpCp_dbb_1816}.

\begin{figure}[!htpb]
\centering
\includegraphics[scale=0.3,angle=270]{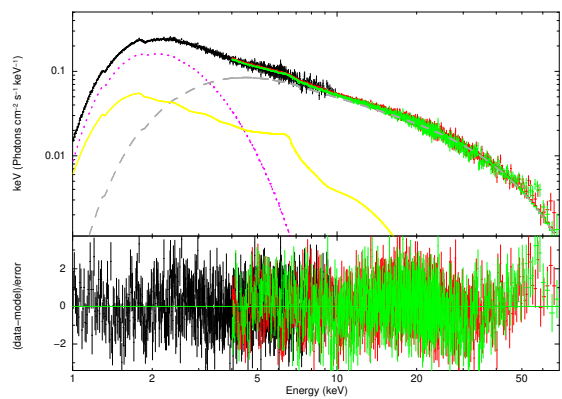}\\
\caption{The best fitting results with the reflection model M5 for NICER/NuSTAR observation of MAXI J1816--195. We show the fitted spectra and the individual model components (main panel), and the residuals in terms of sigmas (sub panel) in the plots. The components $diskbb$, $compTT$, $relxillns$ are plotted with purple dotted line, gray dashed line and yellow solid line, respectively.}
\label{relxill_NS_1816}
\end{figure}

In Table \ref{relxillNS}, we show the results from the fit of model M5. The fitted temperature of the $bbodyrad$ and $diskbb$ components is, $\sim$ 2.32 keV and $\sim$ 0.60 keV, respectively. The temperature of the seed photons for the corona is $0.98 \pm 0.04$ keV, which is in-between the one of the $bbodyrad$ and the $diskbb$, suggesting that the seed photons are likely a mixture of the two components. The electron temperature $kT_{e}$ in the corona is $9.07_{-0.20}^{+0.16}$ keV, with the optical depth being $6.77_{-0.06}^{+0.15}$. The ionization parameter of the disk is similar to that in $relxillCp$ and $relxilllpCp$, log($\xi$) $\sim$ 2. The inner radius of the disk is less than $\sim$ 9.25 $R_{g}$, with the reflection fraction parameter in $relxillns$ being $\sim$ 21. The corresponding spectra, individual components, and residuals of the fits are present in Figure \ref{relxill_NS_1816}.



\begin{table*}[h]
\centering
\caption{Best-fitting results for the fits to the X-ray spectra of 4U 1636--53 with the reflection model M1-M4. We give the unabsorbed flux (erg cm$^{-2}$ s$^{-1}$) in the energy range 0.1-100 keV. All errors in the tables are at the 90 per cent confidence level unless otherwise indicated. A symbol * means that the error pegged at the hard limit of the parameter range.}
\resizebox{\textwidth}{!}{
\begin{tabular}{cccccc}
\hline 
Model Comp        & Parameter                          & M1                        &M2                           &M3                           &M4       \\
\hline                                                                              
{\sc tbabs}       &$N_{\rm H}$ ($10^{22}$ $cm^{-2}$)   &$0.82 \pm 0.04$            &$0.87 \pm 0.03$              &$0.82_{-0.05}^{+0.04}$       &$0.76_{-0.02}^{+0.04}$    \\
                                                                                                                                              
{\sc bbodyrad}    & $kT_{BB}$ (keV)                    &$1.32 \pm 0.07$            &$1.50 \pm 0.02$              &$1.80_{-0.05}^{+0.04}$       &$1.72_{-0.06}^{+0.05}$ \\
                  & Norm                               &$12.0_{-3.3}^{+3.9}$       &$3.2 \pm 0.2$                &$6.0_{-0.5}^{+0.8}$          &$2.8 \pm 0.1$ \\
                  & Flux ($10^{-10}$ c.g.s)            &$3.92_{-0.43}^{+0.41}$              &$1.73\pm 0.05$               &$6.77_{-0.07}^{+0.11}$       &$2.61_{-0.29}^{+0.34}$ \\
                                                                                                                                              
{\sc diskbb}      & $kT_{disk}$(keV)                   &$0.160 \pm 0.002$          &$0.157 \pm 0.002$            &$0.158_{-0.004}^{+0.007}$    &$0.167_{-0.004}^{+0.002}$ \\
                  & Norm                               &$151579_{-26033}^{+22572}$ &$223547_{-34060}^{+39985}$   &$170957_{-62654}^{+59603}$   &$101370_{-26358}^{+30441}$  \\
                  & Flux ($10^{-10}$ c.g.s)            &$18.4_{-2.5}^{+2.8}$       &$24.6_{-2.4}^{+2.5}$         &$19.7_{-3.7}^{+4.0}$         &$14.6_{-2.7}^{+3.0}$ \\
                                                                                                                                              
 {\sc thcomp}     & $\Gamma$                           &$1.68 \pm 0.02$            &$1.65 \pm 0.01$              &$1.50_{-0.01}^{+0.03}$       &$1.53_{-0.02}^{+0.04}$  \\
                  & $kT_{e}$ (keV)                     &$33.0_{-3.7}^{+3.3}$       &$18.2_{-0.7}^{+0.8}$         &$17.8_{-0.6}^{+0.8}$         &$18.0_{-0.7}^{+0.8}$  \\
                  & $f_{cov}$                          &$0.79_{-0.06}^{+0.05}$     &$0.05 \pm 0.01$              &$0.54_{-0.04}^{+0.05}$       &$0.05 \pm 0.01$  \\
		                                                                                                                                      
{\sc relxillCp}   & Incl                               &$15.0_{-10.0*}^{+9.3}$     &$5.0_{-0*}^{+10.1}$          &--                           &--    \\
                  & $R_{\rm in}$ ($R_{\rm g}$)         &$15.7_{-3.7}^{+5.6}$       &$13.1_{-1.8}^{+2.2}$         &--                           &--    \\
	              & log($\xi$)   	                   &$3.94_{-0.04}^{+0.02}$     &$3.90_{-0.01}^{+0.03}$       &--                           &--     \\
	              & Norm ($10^{-4}$)                   &$21.7_{-1.8}^{+2.0}$       &$14.1_{-1.2}^{+1.1}$         &--                           &--     \\
	              & Flux ($10^{-10}$ c.g.s)            &$25.2 \pm 1.7$            &$16.9_{-1.5}^{+1.1}$          &--                           &--   \\
                                                                                                                                              
{\sc relxilllpCp} & Incl                               &--                        &--                            &$34.7_{-1.5}^{+1.6}$         &$18.5_{-8.0}^{+6.7}$ \\
                  & $R_{\rm in}$ ($R_{\rm g}$)         &--                        &--                            &$6.6_{-0.1}^{+3.2}$          &$17.2_{-3.4}^{+4.6}$ \\
                  &  $h$ ($R_{\rm g}$)                 &--                        &--                            &$2.31 \pm 0.01$              &$2.51 \pm 0.05$ \\
                  & log($\xi$)                         &--                        &--                            &$3.59_{-0.02}^{+0.01}$       &$3.78 \pm 0.04$   \\
                  & Norm ($10^{-2}$)                   &--                        &--                            &$9.3_{-2.5}^{+2.8}$          &$3.2_{-0.5}^{+0.6}$   \\
	              & Flux ($10^{-10}$ c.g.s)            &--                        &--                            &$19.8 \pm 0.2$               &$16.4_{-0.9}^{+0.8}$     \\
 \hline                                               
                  & $\chi_{\nu}^{2}$ ($\chi ^{2} / dof$) &$1.23(3476.4/2831)$     &$1.24(3505.9/2831)$          &$1.22(3447.6/2830)$           &$1.22(3443.8/2830)$    \\
                  & Total Flux ($10^{-10}$ c.g.s)        &$54.4_{-2.5}^{+3.2}$    &$59.6_{-3.3}^{+1.8}$         &$54.6_{-4.4}^{+6.2}$          &$49.0_{-2.8}^{+3.1}$     \\
\hline                                                     
\end{tabular}}
\label{4u1636}
\end{table*}

\begin{table*}[h]
\centering
\caption{Best-fitting results for the fits to the X-ray spectra of XTE J1739--285 with the reflection model M1-M4. We give the unabsorbed flux (erg cm$^{-2}$ s$^{-1}$) in the energy range 0.1-100 keV. A symbol * means that the error pegged at the hard limit of the parameter range.}
\resizebox{\textwidth}{!}{
\begin{tabular}{cccccc}
\hline 
Model Comp        & Parameter                      & M1                         &M2                            &M3                           &M4         \\              
\hline                                                                                                                                                                                          
{\sc tbabs}   &$N_{\rm H}$ ($10^{22}$ $cm^{-2}$)   &$2.81_{-0.20}^{+0.19}$      &$3.23_{-0.19}^{+0.13}$        &$3.06_{-0.44}^{+0.16}$       &$3.15_{-0.15}^{+0.14}$   \\
                                                 
{\sc bbodyrad}    & $kT_{BB}$ (keV)                &$0.88_{-0.07}^{+0.10}$      &$1.32_{-0.04}^{+0.05}$        &$1.48_{-0.18}^{+0.27}$       &$1.30_{-0.04}^{+0.09}$   \\
                  & Norm                           &$31.3_{-12.8}^{+14.4}$      &$1.5_{-0.1}^{+0.3}$           &$4.5_{-3.0}^{+1.8}$          &$1.7 \pm 0.3$ \\
                 & Flux ($10^{-10}$ c.g.s)         &$2.0 \pm 0.2$               &$0.5 \pm 0.1$                 &$2.3_{-0.3}^{+0.2}$          &$0.5 \pm 0.1$ \\
                                                
{\sc diskbb}       & $kT_{disk}$ (keV)             &$0.14_{-0.02}^{+0.01}$      &$0.14 \pm 0.01$               &$0.15_{-0.01}^{+0.03}$       &$0.15 \pm 0.01$  \\
                 & Norm                            &$118462_{-68334}^{+186391}$ &$383888_{-153209}^{+276573}$  &$198676_{-120239}^{+227325}$ &$229750_{-109815}^{+191185}$     \\
                 & Flux ($10^{-10}$ c.g.s)         &$9.0_{-4.1}^{+5.8}$         &$26.5_{-7.3}^{+12.2}$         &$15.8_{-12.0}^{+10.6}$       &$19.7_{-7.7}^{+14.0}$     \\

{\sc thcomp}      & $\Gamma$                       &$1.78_{-0.01}^{+0.02}$      &$1.73_{-0.01}^{+0.03}$        &$1.65_{-0.02}^{+0.04}$      &$1.68_{-0.04}^{+0.03}$  \\
                  & $kT_{e}$ (keV)                 &$40.1_{-11.5}^{+7.4}$       &$18.1_{-1.2}^{+1.8}$          &$19.8_{-2.6}^{+3.2}$        &$21.3_{-2.4}^{+4.3}$  \\
                  & $f_{cov}$                      &$1.00_{-0.09}^{+0*}$        &$0.03_{-0.01}^{+0.02}$        &$0.79_{-0.19}^{+0.07}$      &$0.03 \pm 0.01$   \\
                                                   
{\sc relxillCp}     & Incl                         &$29.0_{-24.0*}^{+11.2}$     &$5.0_{-0*}^{+31.1}$           &--                          &--  \\
                 & $R_{\rm in}$ ($R_{\rm g}$)      &$19.9_{-10.8}^{+25.6}$      &$18.0_{-12.6*}^{+10.8}$       &--                          &--   \\
                 & log($\xi$)                      &$3.44_{-0.19}^{+0.18}$      &$3.18_{-0.28}^{+0.32}$        &--                          &--   \\
                 & Norm ($10^{-4}$)                &$3.65_{-0.94}^{+1.24}$      &$1.14_{-0.30}^{+0.53}$        &--                          &--   \\
                 & Flux ($10^{-10}$ c.g.s)         &$4.1_{-0.9}^{+1.0}$         &$1.4_{-0.4}^{+1.2}$           &--                          &--    \\
                                          
{\sc relxilllpCp}   & Incl                         &--                          &--                            &$5.0_{-0*}^{+31.3}$        &$14.0_{-9.0*}^{+29.5}$ \\
                 & $R_{\rm in}$ ($R_{\rm g}$)      &--                          &--                            &$17.7_{-5.1}^{+29.2}$      &$22.4_{-9.6}^{+35.4}$  \\
                 &  $h$ ($R_{\rm g}$)              &--                          &--                            &$2.20_{-0*}^{+0.20}$       &$2.54_{-0.17}^{+0.69}$   \\
                 & log($\xi$)                      &--                          &--                            &$3.57_{-0.11}^{+0.10}$     &$3.56_{-0.89}^{+0.13}$   \\
                 & Norm ($10^{-2}$)                &--                          &--                            &$3.46_{-1.13}^{+0.66}$     &$0.39_{-0.37}^{+0.68}$  \\
                 & Flux ($10^{-10}$ c.g.s)         &--                          &--                            &$4.3 \pm 0.4$              &$2.3_{-1.3}^{+0.4}$ \\ 
\hline                                                                                 
          &$\chi_{\nu}^{2}$  $(\chi ^{2} / dof)$   &$1.02(1641.5/1615)$         &$1.02(1652.7/1615)$           &$1.02(1642.0/1614)$        &$1.02(1649.1/1614)$   \\
          &Total Flux ($10^{-10}$ c.g.s)           &$19.6_{-4.2}^{+2.8}$        &$36.8_{-7.5}^{+11.8}$        &$26.1_{-9.9}^{+11.1}$      &$29.9_{-7.4}^{+13.2}$  \\                                          
\hline                                                                                                                                  
\end{tabular}}                                                                                  
\label{j1739}
\end{table*}

\begin{table*}[h]
\centering
\caption{Best-fitting results for the fits to the X-ray spectra of MAXI J1816--195 with the reflection model M1-M4. We give the unabsorbed flux (erg cm$^{-2}$ s$^{-1}$) in the energy range 0.1-100 keV. A symbol * means that the error pegged at the hard limit of the parameter range.}
\begin{tabular}{cccccc}
\hline 
Model Comp        & Parameter                  &M1                          &M2                          & M3                 	    &  M4                    \\  
\hline                                                                     
{\sc edge}  &$E$ (keV)			               &$1.82_{-0.01}^{+0.02}$      &$1.83 \pm 0.02$             &$1.83 \pm 0.02$           &$1.83 \pm 0.02$   \\
			&$\tau$ ($10^{-2}$)		           &$4.63_{-0.94}^{+0.93}$      &$4.14_{-1.04}^{+1.05}$      &$4.86_{-0.98}^{+1.03}$    &$4.38_{-0.99}^{+1.03}$      \\
\\			                                                                             
{\sc tbabs} &$N_{\rm H}$ ($10^{22}$ $cm^{-2}$) &$2.60 \pm 0.03$             &$2.62 \pm 0.03$             &$2.60_{-0.02}^{+0.03}$    &$2.62_{-0.02}^{+0.03}$    \\                     
			                                                             
{\sc bbodyrad}   & $kT_{BB}$ (keV)             &$1.05 \pm 0.04$             &$1.31 \pm 0.05$             &$1.11_{-0.04}^{+0.05}$    &$1.35_{-0.04}^{+0.05}$      \\
			& Norm                             &$78.3_{-10.2}^{+12.7}$      &$4.5_{-1.2}^{+1.1}$         &$62.4_{-10.0}^{+12.2}$    &$5.1 \pm 0.7$  \\
			& Flux ($10^{-10}$ c.g.s)          &$10.3 \pm 0.3$              &$1.4 \pm 0.3$               &$10.1_{-0.2}^{+0.3}$      &$1.8 \pm 0.3$ \\
	                                          
{\sc diskbb}	& $kT_{disk}$ (keV)	           &$0.56 \pm 0.01$             &$0.52 \pm 0.01$             &$0.57 \pm 0.01$            &$0.53 \pm 0.01$    \\
			& Norm                             &$1312_{-106}^{+112}$        &$1822_{-168}^{+195}$        &$1180_{-100}^{+120}$       &$1583_{-169}^{+183}$     \\
			& Flux ($10^{-10}$ c.g.s)          &$27.0 \pm 0.7$              &$27.6 \pm 0.7$              &$26.4_{-0.5}^{+1.0}$       &$26.8_{-0.4}^{+0.7}$      \\
		                                      
{\sc thcomp}      & $\Gamma$                   &$2.04 \pm 0.02$             &$2.02_{-0.02}^{+0.03}$      &$2.02_{-0.02}^{+0.03}$     &$2.00 \pm 0.02$     \\
			& $kT_{e}$ (keV)                   &$13.6_{-0.7}^{+0.8}$        &$13.3_{-0.7}^{+0.8}$        &$14.8_{-1.0}^{+1.1}$       &$14.7_{-0.9}^{+0.5}$     \\
			&$f_{cov}$                         &$0.80_{-0.03}^{+0.04}$      &$0.30 \pm 0.02$             &$0.77_{-0.03}^{+0.04}$     &$0.27 \pm 0.02$     \\
					           
{\sc relxillCp}   & Incl                       &$16.4_{-5.8}^{+4.3}$        &$14.2_{-6.4}^{+4.4}$        &--                        &--  \\
			& $R_{\rm in}$ ($R_{\rm g}$)       &$5.16_{-0*}^{+0.23}$        &$5.16_{-0*}^{+0.22}$        &--                        &--  \\
			& log($\xi$)                       &$2.19_{-0.07}^{+0.08}$      &$2.24_{-0.07}^{+0.08}$      &--                        &--    \\
			& Norm ($10^{-4}$)                 &$25.1_{-3.2}^{+3.5}$        &$25.3_{-3.0}^{+3.2}$        &--                        &--  \\    
			& Flux ($10^{-10}$ c.g.s)          &$37.5_{-5.1}^{+6.1}$        &$37.2_{-5.5}^{+5.0}$        &--                        &--  \\
                                                                                                                                   
{\sc relxilllpCp}   & Incl                     &--                        &--                            &$21.3_{-3.5}^{+3.7}$       &$20.5_{-3.7}^{+2.9}$   \\ 
			& $R_{\rm in}$ ($R_{\rm g}$)       &--                        &--                            &$5.16_{-0*}^{+0.16}$       &$5.16_{-0*}^{+0.13}$   \\
			&  $h$ ($R_{\rm g}$)               &--                        &--                            &$2.99_{-0.22}^{+0.60}$     &$2.99_{-0.32}^{+0.51}$     \\
			& log($\xi$)                       &--                        &--                            &$2.16_{-0.07}^{+0.08}$     &$2.20_{-0.06}^{+0.08} $     \\ 
			& Norm($10^{-2}$)                  &--                        &--                            &$3.0_{-1.5}^{+3.4}$        &$3.1_{-1.4}^{+2.8}$  \\
			& Flux ($10^{-10}$ c.g.s)   	   &--                        &--                            &$42.1_{-5.1}^{+7.1}$       &$42.6_{-4.8}^{+5.7}$    \\        
\hline                            
      &$\chi_{\nu}^{2}$  $(\chi^{2} / dof)$   &$1.06(2062.2/1945)$        &$1.07(2077.8/1945)$         &$1.05(2041.4/1944)$         &$1.05(2050.3/1944)$  \\
      &Total Flux ($10^{-10}$ c.g.s)          &$85.3_{-4.7}^{+5.2}$       &$85.2_{-4.3}^{+4.8}$        &$88.9_{-4.1}^{+6.0}$        &$88.9_{-4.1}^{+4.9}$   \\
			
\hline                                  
\end{tabular}
\label{j1816}
\end{table*}

\begin{table}[h]
\centering
\caption{Best-fitting results for the fit to the X-ray spectra of MAXI J1816--195 with the reflection model M5. We give the unabsorbed flux (erg cm$^{-2}$ s$^{-1}$) in the energy range 0.1-100 keV. A symbol * means that the error pegged at the hard limit of the parameter range.}
\begin{tabular}{ccc}
\hline 
Model Comp        & Parameter                    & M5    \\
\hline
{\sc edge}  &$E$ (keV)							 &$1.81 \pm 0.02$  \\
			& $\tau$ ($10^{-2}$)	   			 &$3.78_{-0.98}^{+0.88}$       \\
\\
{\sc tbabs} &$N_{\rm H}$ ($10^{22}$ $cm^{-2}$)   &$2.58_{-0.02}^{+0.03}$       \\
			                                        
{\sc diskbb} &$kT_{disk}$ (keV)          	     &$ 0.60_{-0.01}^{+0.03}$        \\
			&Norm                                &$935_{-163}^{+47}$     \\
			& Flux ($10^{-10}$ c.g.s)            &$24.9 \pm 0.5$     \\
                                                    
{\sc compTT}  &$T_{0}$  (keV)                    &$0.98 \pm 0.04$   \\
			&$kT_{e}$ (keV)                      &$9.07_{-0.20}^{+0.16}$   \\
			&$\tau$                              &$6.77_{-0.06}^{+0.15}$     \\
			&Norm ($10^{-3}$)                    &$24.9_{-0.7}^{+0.5}$     \\
			&Flux ($10^{-10}$ c.g.s)             &$21.8_{-0.4}^{+0.3}$    \\

{\sc relxillNS}	&Incl                            &$25.4_{-3.4}^{+4.0}$     \\
		   	&$R_{\rm in}$ ($R_{\rm g}$)          &$8.00_{-0.81}^{+1.25}$     \\
			& $kT_{BB}$ (keV)                    &$2.32_{-0.03}^{+0.01}$     \\
			&log($\xi$)                          &$2.00_{-0.03}^{+0.05}$     \\
			&$f_{refl \_ ns}$                    &$21.08_{-1.69}^{+10.31}$     \\
			&Norm ($10^{-5}$)                    &$1.19_{-0.19}^{+0.13}$     \\
			& Flux ($10^{-10}$ c.g.s)            &$31.7_{-1.7}^{+2.4}$     \\
\hline                                              
			&$\chi^2_\nu$ ($\chi^2/dof)$         &$1.16(2250.6/1944)$     \\
			&Total Flux ($10^{-10}$ c.g.s)       &$77.7_{-1.7}^{+3.5}$     \\
\hline                                              
\end{tabular}
\label{relxillNS}
\end{table}

\section{Discussion} 

In this work, we investigated the three neutron star low-mass X-ray binaries with simultaneous NICER plus NuSTAR observations. We found that for the systems in the hard spectral state, if seed photons come from the NS, more than $\sim$ 50\% of the NS photons go into the corona; While if the seed photons come from the accretion disk, only $\sim$ 3\%-5\% disk photons move into the corona. Furthermore, the height of the corona is very small, $\sim$ 2-3 $R_{g}$ from the central compact object. More importantly, we found that the seed photon type has significant influence in the inferred neutron star radiation.

\subsection{Boundary layer and magnetic filed in 4U 1636--53 and XTE J1739--285}
The derived inner disk radius of 4U 1636--53 and XTE J1739--285 is larger than the typical NS radius, suggesting that the disk may be truncated by certain mechanism. We thus first estimated the radius of the boundary layer from the mass accretion rate, based on the formula below \citep{popham01},

\begin{equation}
\text{log}(R_\text{max}-R_\text{NS})\simeq5.02+0.245\left|\text{log}\left(\frac{\dot{m}}{10^{-9.85}\:M_{\odot}\:\text{yr}^{-1}}\right)\right|^{2.19}
\end{equation}

\noindent with the mass accretion rate, $\dot{m}$, calculated using the below equation \citep{Galloway2008-Dis,girid24},

\begin{equation}
\dot{m}=\frac{ 6.7 \times 10^{3} \times F_\text{bol} (1+z) (D_{10})^{2}} {M_\text{NS} \times R_\text{NS}}
\end{equation}

\noindent where $F_\text{bol}$ is the total bolometric flux in units of 10$^{-9}$ erg cm$^{-2}$ s$^{-1}$, $z$ is the surface redshift, and $D_{10}$ is the distance of the source in units of 10 kpc. We used a distance of 6 kpc \citep{Galloway2006-1636-6kpc} and 4 kpc \citep{Bailer-Jones-2018-4u1735} for 4U 1636--53 and XTE J1739--285, respectively. We took $F_\text{bol}$ as the total unabsorbed flux in 0.1-100 keV in the fits, and calculated $1+z=(1-\frac{2GM_{NS}}{R_{NS}c^{2}})^{-\frac{1}{2}}$ \citep{Thomas_1+z} where $M_{NS}$=1.4 $M_{\odot}$ and $R_{NS}$=10 km. The derived maximum boundary layer radius is $\sim$ 6-7 $R_{g}$ and $\sim$ 5 $R_{g}$ for 4U 1636--53 and XTE J1739--285, respectively.

Assuming that the accretion disk is truncated by magnetic field for 4U 1636--53 and XTE J1739--285 in hard state, we could place an upper limit on the strength of the field using the following equation \citep{cackett09}, $\mu=3.5 \times 10^{23} k_{A}^{-7/4} x^{7/4} (\frac{M}{1.4M_{ \odot}})^{2} (\frac{f_{ang}}{\eta})^{1/2} (\frac{F_{bol}}{10^{-9} erg/s/cm^{2}})^{1/2} \frac{D}{3.5 kpc}$, where $\mu$ is the magnetic dipole moment, $k_{A}$ is the coefficient that depends on the conversion from spherical to disc accretion, $x$ is given from $R_{in}$=$x$ $GM/c^{2}$ \citep{cackett09}, $f_{ang}$ is the anisotropy correction factor and $\eta$ is the accretion efficiency. 

We then calculated the magnetic field of the sources based on $B$=$\mu/R^{3}$ \citep{Long2005-magnetic}, where $R$ is the radius of the neutron star. During the calculation, we set $k_{A}$ = 1, $f_{ang}$ = 1, $\eta$ = 0.1, $M$=1.4 $M_{ \odot}$, $R$=10 km, and $F_{bol}$=$F_{0.1-100 keV}$. We take the $R_{in}$ and $F_{bol}$ from the fits with M1 model and get the polar magnetic field of 4U 1636--53 and XTE J1739--285 as B $\le$ $1.9\times10^{9}$ G and B $\le$ $ 2.8\times10^{9}$ G, respectively. 

The calculated upper limit of the magnetic field in XTE J1739--285 in this work is a little bigger than the one (B $\le$ $6.2 \times 10^{8} $ G) in the work of \citet{mondal22}. For 4U 1636--53, \citet{stroh99} proposed that there likely be a stronger magnetic field in this source than others according to the fact that 4U 1636--53 is the only source showing significant pulsations at the sub-harmonic of the strongest oscillation frequency. \citet{mondal21} estimated the magnetic field strength of 4U 1636--53 at the poles as $B < 2.0 \times10^{9}$ G, close to the one that derived in this work. So both the magnetic field measured in \citet{mondal21} and in this work reside between the typical range of 10$^{8}$ G to 10$^{9}$ G \citep[e.g.,][]{mukh15,lud16,lud17,mon19}, showing no clear indication that 4U 1636--53 has stronger magnetic field than others.

\subsection{Corona geometry}
In the current stage, the physical and geometrical properties of the corona are still unclear. Three typical types of the corona model have been proposed: (1) static corona with different geometries \citep[e.g., slab;][]{haardt94}; (2) base of a jet \citep[][]{markoff01,markoff05}; (3) radiatively inefficient accretion flow \citep[RIAF;][]{ferreira06,narayan08}. Besides, it has also been proposed that the boundary layer likely plays the role of the corona in the neutron star systems \citep[e.g.,][]{popham01,dai10,saavedra23}. 

The cover fraction of the seed photons is important for the understanding of the position and the geometry of the corona. The fraction derived in this work favors those corona models in which the corona is close to the central compact object in the hard state: in this case, many photons from the NS could go into the nearby corona, thus the corresponding cover fraction is high (larger than $\sim$ 50\% in our results); while the disk is far from the neutron star in the hard state, so the fraction of the disk photons entering into the corona is low (only $\sim$ 3\%-5\% in our fits). This scenario is also supported by the derived small height of the corona in our fits.

The above results support both the lamp-post geometry and the boundary layer scenario. In the well-known lamp-post case, the corona could be the base of a jet, or a moving hot plasma situated above the spin axis of the central compact object. when the height of the corona is small, large fraction of the thermal photons from NS surface could get into the corona and be Compton scattered, consistent with the big cover fraction derived in this work. The boundary layer is the region where the in-falling matter decelerates from its orbital velocity in the accretion disk to the rotation velocity of the neutron-star surface \citep{ss86,Inoga99}. \citet{popham01} found that the main part of the boundary layer is hot ($\ge$10$^{8}$ K), and is vertically and radially extended. For LMXBs in the low hard state, Comptonization in the hot boundary layer is able to generate a power-law spectrum with a cutoff energy of $\sim$ 20-30 keV. Observationally, \citet{wyn17} derived a small height of the corona ($\sim$ 2-3 $R_{g}$) in 4U 1636--53 with NuSTAR observations and hence proposed that the hard X-rays likely come from the boundary layer. Similarly, \citet{Thompson05} applied two comptonization components for the spectroscopy of the GS 1826-238, and one of them is proposed to come from, or near, the boundary layer. Interestingly, the derived maximum boundary layer radius ($\sim$ 5-7 $R_{g}$) in this work is close to the height of the corona ($\sim$ 3 $R_{g}$). So it is possible that the boundary layer plays the role of the corona in the two hard observations in this work. 

One way to distinguish a moving corona/base of a jet from a boundary layer is to see whether the corona has big velocity: For a moving corona/jet base, they should have mildly relativistic velocities \citep{Beloborodov99,king17}; While in the case of a boundary layer, \citet{popham01} show that the radial velocity is well below the sound speed in their calculations. Here we could not constrain the velocity of the corona in the component $relxilllpCp$. Observations with long enough exposure time are required in the future to separate the lamp-post geometry and the boundary layer out for the NS systems.

\begin{table*}[h]
\renewcommand{\arraystretch}{0.8}
\centering
\caption{Details of the NS LMXB samples collected from literatures for Figure \ref{l-kT}. }
\begin{threeparttable}
\tiny
\begin{tabular}{ccccccccc}
\hline
  Number &Source        &Telescope        & $kTe$ (keV)             &$E_{cut}$ (keV)                   &Flux  $^{a}$         &D (kpc)   & References \\
\hline                                                                                                                                                                                  
 1  & GX 3+1        & NuSTAR             &$2.45 \pm 0.04$              & --                         & $79.66 \pm 13.48 $      &6.5       & \cite{Ludlam-2019-hard-state} \\
 2 &4U 1702--429    & NuSTAR              & --                          &$53_{-2}^{+11}$            & $8.56 \pm 2.07 $        &5.65      & \cite{Ludlam-2019-hard-state} \\
 3 &4U 0614+091     & NuSTAR              & --                          &$16_{-4}^{+1}$             & $68.75 \pm 14.06$       &3.2       &  \cite{Ludlam-2019-hard-state}  \\
 4 &4U 1746--371    & NuSTAR               &$2.91 \pm 0.04$             &--                         & $3.67 \pm 0.54$         &11.9      &  \cite{Ludlam-2019-hard-state}  \\
                                    
 5 &4U 1636--53     & NuSTAR              & --                         &$135.9 \pm 0.7$             &$20.86  \pm 0.18$        &6         &\cite{wyn17}\\
                                    
 6 &XTE J1710--281  & Suzaku             &$4.45_{-0.55}^{+0.69}$      & --                          &$1.86 \pm 0.06 $         & 15       & \cite{Sharma-2020-hard-state} \\  
 7 &4U 1728--34     &Suzaku              & --                          &$16.6 \pm 1.2$              &$73.6_{-3.0}^{+1.5}$     &5$^{*}$   &\cite{Lyu-2019-MNRAS}  \\
 8  &4U 1728--34    &XMM-Newton/RXTE     & --                        &$19.95_{-2.00}^{+7.03}$       &$222_{-19}^{+66}$       &5$^{*}$    & \cite{Lyu-2019-MNRAS} \\
\hline
\end{tabular}

\footnotesize
$^{a}$  Flux is in unit of $10^{-10} erg/s/cm^{2}$. We used the total unabsorbed flux when the power-law flux is not available in the literatures. The flux collected from the literatures are scaled to the one in a unit energy range 0.1-100 keV for the following compactness calculation. $^{*}$ The distance is taken from \cite{Vincentelli2023-1728-5kpc}, which is an approximation of the result in the work of \cite{Galloway2003-1728-5kpc}.


\end{threeparttable}
\label{l-kT-da}
\end{table*}

\subsection{Corona compactness}
\begin{figure}[!htpb]
\centering
\includegraphics[scale=0.5]{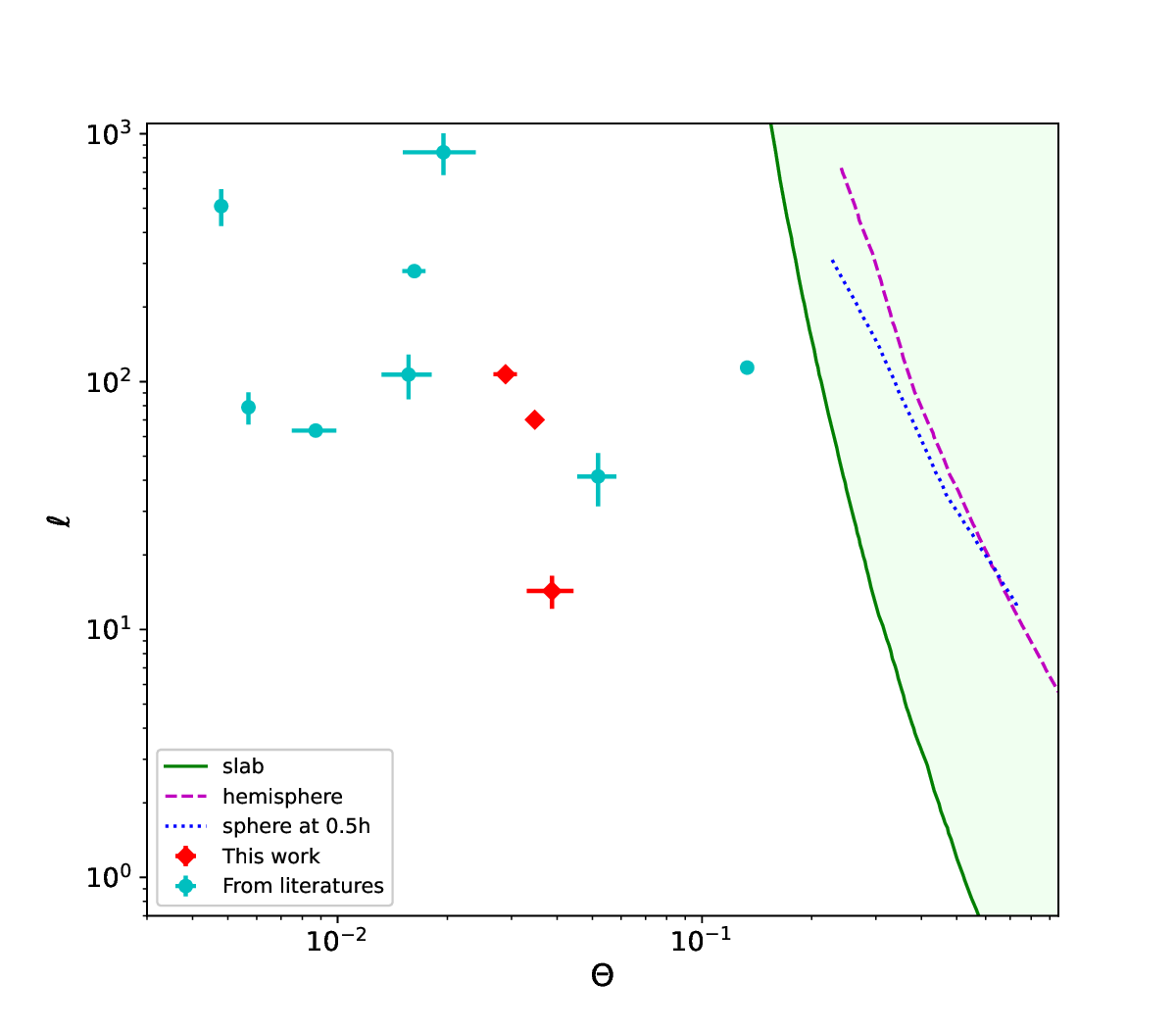}\\
\caption{The compactness–temperature ($\ell$--$\theta$) diagram for the NS LMXBs. We used different symbols to mark the samples in this work and in the literatures (Table \ref{l-kT-da}). The compactness is calculated as $\ell$=$4\pi \frac{m_{p}}{m_{e}} \frac{R_{g}}{R} \frac{L}{L_{E}}$ \citep{fabian15}, where $m_{p}$ and $m_{e}$ is the mass of proton and electron, and $R$ and $L$ is the corona radius and luminosity. In the calculation, corona size $R$ is assumed to be 10 $R_{g}$ \citep{fabian15}. The parameter $\theta$ is defined as $\theta=kT_{e}/m_{e}c^{2}$, where $kT_{e}$ is estimated as $kT_{e}$=$E_{cut}$/2 when the $kT_{e}$ is no available in the literatures. For the three sources analyzed in this work, the corona luminosity $L$ was derived from the corona flux in 0.1-100 keV, and the $kT_{e}$ was taken directly from the fits with the $relxilllpCp$ model. In the figure, we also over-plot the pair run-away lines for the three geometries \citep{stern95} and show the runaway pair-production forbidden region (green area) for the slab geometry.
}
\label{l-kT}
\end{figure}

In this work, we also studied the physics of the corona for NS systems with the so-called compactness ($\ell$--$\theta$) diagram (Figure \ref{l-kT}). The $\ell$--$\theta$ diagram has already been proved to be a useful tool for understanding the physics inside the corona \citep[e.g.,][]{fabian15,fabian17,Zdzia21,kang22,Berto22,Kamraj22,Tortosa22,hinkle21,kang21,Reeves21}. In the $\ell$--$\theta$ diagram, the dimensionless compactness parameter $\ell$ \citep{guilbert83} represents the ratio of the corona luminosity to its size, and the $\theta$ is proportional to the electron temperature. Physically, when there is power fed into the corona, temperature of the electrons increases thus the scattering of the soft seed photons generate more energetic power-law photons. As the power-law spectrum extends to a Wien tail above $\sim$ 2$m_{e}c^{2}$, the photon-photon collisions could create electron-positron pairs. In this case, new injection of energy into the corona could not lead to the increase of the corona temperature but instead results in an increase of the number of the pairs. So there should be a pair-production limit in the $\ell$--$\theta$ diagram if the corona is pair-dominated. 

\citet{fabian15} showed that the corona in AGNs are hot and radiatively compact, lying close to the pair run-away line in the $\ell$--$\theta$ diagram. This finding suggests that the temperature of the corona in these AGNs is perhaps being regulated by pair production and annihilation. Later, \citet{fabian17} proposed that the corona lying well below the pair limit could also be pair dominated if the corona contains non-thermal electrons, which is able to decreases the pair production threshold temperature.

Interestingly, corona lying beyond the predicted pair run-away line has also been reported in both the AGNs and the black-hole binary (BHB) system. \citet{kang22} recently found that a considerable fraction of AGNs lie beyond the boundaries of the forbidden areas due to runaway pair production. They proposed that the corona likely be more extended than 10 $R_{g}$ in these systems. Furthermore, for the accreting black hole binary XTE J1752-223 in the hard state, \citet{Zdzia21} found that the compactness is much above the maximum allowed by pair equilibrium. The reason behind this discrepancy is still unclear.

The compactness of the corona in the neutron star binaries gets far less attention compared with those in the black hole systems. Here, for the first time, we show the $\ell$--$\theta$ diagram for 9 NS systems (including samples taken from literature listed in Table \ref{l-kT-da}) in Figure \ref{l-kT}. Apparently, all the NS systems lie to the left side of the pair-production forbidden region, away from the slab pair line. This result suggests that the corona in these NS systems are likely not composed of electron-positron pairs, which could be explained by the additional cooling by the thermal seed photons from the neutron star surface. With this cooling process, the temperature of the electrons in the corona is relative low, consequently, the power-law photons generated from the Compton scattering would not be energetic enough to create electron-positron pairs in the following photon-photon collisions, so the corona in the NS systems would not be pair-dominated. If the above scenario is true, then it constitutes another evidence that the seed photons from the NS surface is crucial for the physics of the neutron star systems. 

However, we can not rule out the possibility that the corona is a hybrid plasma, containing both thermal and non-thermal particles. In this case, the corona with cooler temperatures may still be regulated by the pair production process since its threshold temperature is decreased by the non-thermal electrons, the same as being described in the work of \citet{fabian17}. 

\subsection{Spectral properties of MAXI J1816--195}


\citet{lipp23} studied the NICER plus the NuSTAR observation of MAXI J1816--195 and found that the thermal radiation from the neutron star surface is not significant, with the blackbody temperature $\sim$ 2.05 keV and the normalization $N_{bb}$ < 2. This scenario is consistent with the weak blackbody component in our fitting result when the seed photons come from the accretion disk (M2, M4 in table \ref{j1816}). However, the blackbody component is not weak if the seed photons are supplied by the neutron star surface (M1, M3 in table \ref{j1816}), indicating that the source of the seed photons could greatly influence the significance of the neutron star radiation. This is natural since in the latter case the neutron star needs to provide additional photons for the Compton scattering, so a stronger blackbody component is required. Furthermore, the seed photon type also influences the derived radius of the emitting region on the NS surface, as the radius is directly linked to the parameter $N_{bb}$. Therefore, determining the source of the seed photons is a pre-requisite for accurately estimate the emitting region on the neutron star surface.

\citet{mandal23} applied the model $diskbb+nthcomp$ to the pre-burst spectra of MAXI J1816--195 from the same NICER and NuSTAR observation used in this work. They found that the corresponding electron temperature is relatively low, $\sim$11 keV, thus they deduced that the source is in the soft spectra state. This results is interesting since, most of the AMXPs are typically observed in hard spectral states, with SAX J1748.9--2021 \citep{pintore16} and SAX J1808.4--3658 \citep{di19} as perhaps the only two exceptions reported in the soft state. For AMXPs in the hard state, the electron temperature is generally high, $\sim$ tens of keV \citep[e.g.,][]{marek05,sanna17,sanna18,rai19,disalvo22}, and the NS thermal emission is usually found with relatively low temperatures, i.e. below 1 keV \citep[e.g.,][]{sanna17,rai19,sanna22,marino22}. Here we applied the latest reflection models and obtained a relative high blackbody temperature ($\sim$ 1-1.4 keV) plus a low electron temperature $\sim$ 13-15 keV. More importantly, the derived inner radius of the disk from the reflection component is small, $\sim$ $R_{ISCO}$, consistent with the previously measured radius, $R_{in}$=1.04-1.23 $R_{ISCO}$ in the work of \citet{lipp23}. Thus, the relative high blackbody temperature, plus the low electron temperature and the very small inner radius together suggest that MAXI J1816--195 in this observation is likely not in the hard state, instead, it may be in the soft state or a transitional state.

\section{Conclusion}
Benefited from the wide energy range covered by the NICER detector and the NuSTAR satellite, we explored the spectral properties of the three NS LMXBs with the latest reflection models, with different types of the seed photons being selected. The main conclusions in this work could be briefly summarized as below: 

(1) We found that, for the observation of 4U 1636--53 and XTE J1739--285 in this work, more than $\sim$ 50\% neutron star photons enter into the corona if the neutron star provides seed photons; while if the disk becomes the provider, then only $\sim$ 3\%-5\% disk photons could go to the corona. The fraction for the observation of MAXI J1816--195 is a bit higher, $\sim$ 80\% and $\sim$ 30\% in the case of the NS seed photons and the disk seed photons, respectively. These results suggest that the corona should be close to the central neutron star, favoring the lamp-post geometry or the boundary layer scenario.

(2) The derived height of the corona in the lamp-post geometry is relatively low, $\sim$ 3 $R_{g}$, consistent with the scenario that the corona is in the vicinity of the neutron star.

(3) We found that the source of the seed photons have big influence in the significance of the neutron star radiation.  

(4) The compactness diagram of the NS systems reveals that all NS systems investigated in this work reside in the left side of the forbidden pair line, indicating that either the corona is not pair-dominated, or there exists non-thermal electrons inside the corona.   

(5) The derived parameters from the fits suggests that MAXI J1816--195 in this work is likely in a soft or transitional spectral state, rather than the typical hard state of the AMXPs.

\begin{acknowledgements}
We thank the anonymous referee for his/her careful reading of the manuscript and useful comments and suggestions. This research has made use of data obtained from the High Energy Astrophysics Science Archive Research Center (HEASARC), provided by NASA's Goddard Space Flight Center. This research made use of NASA's Astrophysics Data System. Z.Y.F. is grateful for the support from the Postgraduate Scientific Research Innovation Project of Hunan Province (grant No. CX20220662) and the Postgraduate Scientific Research Innovation Project of Xiangtan University (grant No. XDCX2022Y072). Lyu is supported by Hunan Education Department Foundation (grant No. 21A0096). GB acknowledges the science research grants from the China Manned Space Project. X. J. Yang is supported by the National Natural Science Foundation of China (NSFC 12122302 and 12333005). FG is a CONICET researcher. FG acknowledges support by PIBAA 1275 (CONICET). FG was also supported by grant PID2022-136828NB-C42 funded by the Spanish MCIN/AEI/ 10.13039/501100011033 and “ERDF A way of making Europe”.
\end{acknowledgements}

\bibliographystyle{aa}
\bibliography{biblio}
%

\clearpage

\begin{appendix}
\section{A comparison with the results from the $kT_{\rm BB}$-corrected model NTHRATIO}

As $relxill$ reflection models were originally built to model AGN spectra, where the temperature of the soft-photon source is expected to be rather low, both the $relxill$ and $relxillCp$ model variants are based on a power-law illuminating spectrum, with a fixed low-energy cut-off at 0.01 or 0.05 keV. Such energies are much lower than the temperature expected for a neutron star system, either if the source of soft-photons is originated in the accretion disk or in the neutron star surface. In order to check whether this inconsistency has significant influence in the conclusions of this work, we further applied two models A1 and A2: $Tbabs\times(diskbb+thcomp \times bbodyrad+relconv \times (xillvercp \times nthratio))$ and $Tbabs\times(bbodyrad+thcomp \times diskbb+relconv \times (xillvercp \times nthratio))$, where the $nthratio$\footnote{https://github.com/garciafederico/nthratio} multiplicative model was developed to apply a first-order correction to the soft-excess introduced by the low energy cutoff from the xillver-based models. 

The parameters in the $nthratio$ are the power-law index, $\Gamma$, the electron temperature, $kT_{e}$; and the seed photon temperature, $kT_{seed}$. We linked the $\Gamma$ and the $kT_{e}$ to the ones in $thcomp$, and linked the $kT_{seed}$ to the temperature in either the $bbodyrad$ component (in the case of A1) or the $diskbb$ component (in the case of A2). Thus, by incorporating the $nthratio$ correction, no additional free parameters are invoked. In the fits, we fixed the reflection fraction in $xillverCp$ to -1 so that only the reflection is returned.

As shown in Table \ref{appendix-1816nthratio}, the fitted parameters with the model A1 and A2 are consistent with the ones with the model M1 and M2 in table \ref{j1816}. The cover fraction parameter $f_{cov}$ is $\sim$ 0.80 (M1) and $\sim$ 0.82 (A1) in the case of blackbody seed photons, and it is $\sim$ 0.30 (M2) and $\sim$ 0.31 (A2) in the disk seed photon case. The blackbody temperature in the model M1 and A1 is $\sim$ 1.05 keV and $\sim$ 1.03 keV, and the temperature is 1.31 keV in M2 and 1.32 keV in A2. And there is some difference between the inferred absorption column values, which decreases from $\sim$ 2.6 $\times$ 10$^{22}$~cm$^{-2}$ to a smaller value ($\sim$ 2.42 $\times$ 10$^{22}$~cm$^{-2}$ or $\sim$ 2.54 $\times$ 10$^{22}$~cm$^{-2}$) when $ nthratio$ is used. This is of course expected as a consequence of the soft-excess correction introduced by this multiplicative model. The differences between the fits are very tiny and hence do not change the conclusions in this work. And this is similar to the situation in the work of \citet{lyu23}, in which the influence of the low temperature seed photons in 4U 1636--53 has been studied, using $nthratio$ too. They found that the parameters are approximately the same for the observations in the hard spectral state, and only limited difference is present in the soft observations.

\begin{table}[h]
\centering
\caption{Best-fitting results for the fit to the X-ray spectra of MAXI J1816--195 with the reflection model A1 and A2. A symbol * means that the error pegged at the hard limit of the parameter range.}	
\resizebox{9cm}{!}{
\begin{tabular}{cccc}
\hline   
Model Comp       & Parameter                     &A1                         & A2     \\
\hline   
{\sc edge}  &$E$ (keV)			                &$1.83 \pm 0.02$             &$1.83 \pm 0.02$ \\
            &$\tau$ ($10^{-2}$)	                &$5.70 \pm 0.99$             &$4.03 \pm 1.04 $ \\
\\
{\sc tbabs} &$N_{\rm H}$ ($10^{22}$ $cm^{-2}$)   &$2.42 \pm 0.02$             &$2.54_{-0.02}^{+0.03}$\\

{\sc bbodyrad}	& $kT_{BB}$  (keV)	           &$1.03_{-0.05}^{+0.04}$       &$1.32 \pm 0.05$  \\
                & Norm                          &$83.5_{-12.0}^{+17.4}$       &$4.2_{-1.2}^{+1.1}$  \\
 
{\sc diskbb}   &$kT_{disk}$ (keV)               &$0.56 \pm 0.01$            &$0.52 \pm 0.01$   \\
                 & Norm                         &$1419_{-112}^{+136}$       & $1779_{-171}^{+192}$  \\

{\sc thcomp}     & $\Gamma$                     &$2.06 \pm 0.03$             &$2.02_{-0.02}^{+0.03}$    \\       
                 & $kT_{e}$ (keV)               &$14.6_{-0.9}^{+1.1}$        & $13.7_{-0.8}^{+0.9}$   \\
                &$f_{cov}$                      &$0.82_{-0.04}^{+0.05}$      & $0.31_{-0.02}^{+0.03}$    \\

{\sc  relconv}   & Incl                         &$14.5_{-6.4}^{+4.6}$         &$14.2_{-6.4}^{+4.4}$   \\         
               & $R_{\rm in}$ ($R_{\rm g}$)    &$5.16_{-0*}^{+0.23}$        &$5.16_{-0*}^{+0.21}$   \\

{\sc xillverCp}  & log($\xi$)                    &$2.22_{-0.07}^{+0.09}$     & $2.24_{-0.07}^{+0.08}$ \\
                 & Norm ($10^{-4}$)              &$27.2_{-3.6}^{+4.0}$      & $25.5_{-3.1}^{+3.3}$    \\
\hline                                                   
&$\chi^2_\nu$ ($\chi^2/dof)$                     &$1.06(2054.1/1945)$        &$1.07(2084.8/1945)$       \\
\hline                                                   
\end{tabular}}
\label{appendix-1816nthratio}
\end{table}

\end{appendix}

\end{document}